\documentclass{easychair}

\usepackage{doc}
\usepackage{caption}
\usepackage[export]{adjustbox}
\usepackage{subcaption}

\title{InTAS - The Ingolstadt Traffic Scenario for SUMO}

\author{
Silas C. Lobo\inst{1,2}
\and
    Stefan Neumeier\inst{3}
\and
    Evelio M. G. Fernandez\inst{4}
\and
     Christian Facchi\inst{5}
}

\institute{
   Universidade Federal do Paraná,
   Curitiba, Brazil\\
\and
  Technische Hochschule Ingolstadt,
  Ingolstadt, Germany\\
  \email{silascorreia.lobo@carissma.eu}\\
\and
   Technische Hochschule Ingolstadt,
   Ingolstadt, Germany\\
   \email{stefan.neumeier@thi.de}\\
\and
   Universidade Federal do Paraná,
   Curitiba, Brazil\\
   \email{evelio@eletrica.ufpr.br}\\
\and
   Technische Hochschule Ingolstadt,
   Ingolstadt, Germany\\
   \email{christian.facchi@thi.de}\\
 }
 
\authorrunning{Lobo, Neumeier, Fernandez and Facchi}

\titlerunning{Ingolstadt Traffic Scenario (InTAS)}

\begin{document}

\maketitle

\begin{abstract}
Vehicular Ad Hoc Networks (VANETs) are expected to be the next big step towards safer road transport, supporting applications to exchange information between vehicles. To develop novel applications for this area a high number of tests, considering all traffic situations, are demanded. However, it is unfeasible to reproduce these tests in real life, by the fact that any failure on the applications would cause severe impacts on transport system safety and could risk human lives. Thus, this paper presents the concept, model, and validation for InTAS, a realistic traffic scenario for Ingolstadt. InTAS’ road topology accurately represents Ingolstadt’s real road map. Elements such as buildings, bus stops, and traffic lights were added to the map. Twenty traffic lights systems were simulated according to the real program deployed on the traffic lights. Traffic demand was modeled based on the \texttt{activitygen} method, considering demographic data and real-traffic information. The city’s public transport system was also simulated accordingly to bus time-tables and their routes. The simulation step was implemented considering the best value for \texttt{device.rerouting.probability}, which was defined by evaluating InTAS’ output and real traffic data. The scenario was validated by comparing real-traffic data from 24 measurement points with InTAS’ simulation results.
\end{abstract}

\section{Introduction}
\label{sect:introduction}
Vehicular Ad-hoc Networks (VANETs) have emerged as one of the most promising automotive technologies in the last years. In a near-future, it is expected that VANETs will be a key enabling technology for autonomous driving, road safety, hazard information service, improvement of traffic congestion issues and many other applications \cite{environment}. VANETs are cooperative vehicular networks based on wireless communication among vehicle-to-vehicle (V2V), vehicle-to-infrastructure (V2I), also known as Road Side Unit (RSU), and vehicle-to-everything (V2X). Among these most apparent advances provided by VANETs, the capacity to share pertinent information on-road situations with other vehicles and RSU in real-time must be highlighted. This allows for a better and safer driving experience.
VANET applications are mostly concentrated in three branches: infotainment, traffic management, and safety~\cite{environment}. Infotainment is related to drivers' and passengers' convenience, entertainment, and their relationship with the vehicle. In contrast, traffic management applications focus on the vehicle’s behavior, as soon as it enters the street network. According to the safety branch, VANETs are prepared to deal with different events and then warn the driver about an incident, as adverse weather~\cite{etsi1}, dangerous situation~\cite{etsi2}, impact reduction~\cite{etsi3}, a stationary vehicle~\cite{etsi4}, and others. Dealing with the development of novel VANET applications, especially when focusing on safety, demands a huge amount of tests, e.g. for analyzing and validation. Furthermore, these tests have to consider all possible situations by the fact that any failure on the applications would cause severe impacts on transport system safety and could risk human lives \cite{environment}. Moreover, testing technology like VANET in the real world is not only extremely expensive, but reproducing test cases is hardly possible. Thus, simulation tools are vital to make testing affordable and more reproducible.
Complete VANET simulations require a simulation framework, containing at least a vehicle traffic scenario and a wireless communication protocol implementation \cite{Car2XHIL}. The more realistic the traffic scenario is, the better the simulation results are. A realistic traffic scenario comprises a real-world road topology including all public road categories, ranging from residential streets to highways. This is required to model accurate mobility over a realistic traffic demand. Traffic scenarios have been developed for the transportation community for a while, but their scope is not mainly on evaluating VANET simulations, which demands a deeper view of the mobility patterns, analyzing vehicle’s position and driver’s behavior, known as a microscopic view. Furthermore, city structures as buildings, bridges, and passages have to be emphasized, as they might have an impact on the communication \cite{Buildings}.
Up to now, to the best of our knowledge, there are three freely available realistic traffic scenarios for Simulation of Urban Mobility (SUMO): TAPAS Cologne \cite{TAPAS}, Luxembourg SUMO Traffic \cite{LUST}, and Monaco SUMO Traffic \cite{MOST}.None of the previously mentioned scenarios have modeled a city with characteristics presented in Ingolstadt. This is because the city has peculiarities as a large industry that concentrates approximately half of work positions and operates 24 hours a day in shift operations. The city also detains a high income per inhabitant and an extremely low unemployment rate~\cite{INGOLSTADT}. Thus, using Ingolstadt for a new traffic scenario, the research community can benefit from a new and partly different type of map.
Ingolstadt is located in the state of Bavaria, southern Germany, with an area of 133.36~km\textsuperscript{2} and a registered population of about 135,000 inhabitants counted in December 2018~\cite{INGOLSTADT}. It is ranked the fifth biggest city in the state and has characteristics that extremely influence its traffic, e.g. it is the site of Audi AG, which employs approximately 44,000 workers, representing more than 43\% of Ingolstadt’s work positions~\cite{INGOLSTADT}. Additionally, the car usage rate compared to other cities in Germany is relatively high \cite{verkehrsentwicklungsplan} and a high penetration rate of new cars among the inhabitants is noticed. Additionally, in the near-future the city will implement a Car2X communication system~\cite{audi-vernetzt}.
Thus, this work's proposal is to develop a realistic traffic scenario enclosing Ingolstadt city and enable it especially for the use case of V2X simulations, considering traffic flow and driver’s behavior. This will help to speed up the development of Car2X systems. Furthermore, such a simulation can be used in other systems such as advanced driving simulators like OpenROUTS3D \cite{OpenROUTS3D}. Moreover, this scenario will be developed using SUMO, which is a powerful Open-Source simulator that supports large road networks. It is suitable for both macroscopic and microscopic simulations and provides great interaction with network simulators such as Omnet++ \cite{SUMO}.
To introduce the Ingolstadt City scenario, this paper is structured as follows. Chapter \ref{sect:relatedworks} presents related work. Chapter \ref{sect:InTAS} introduces the Ingolstadt Traffic Scenario (InTAS) by explaining concept and model. Subsequently, Chapter \ref{sect:evaluation} presents the scenario's validation. Finally, Chapter \ref{sect:future-work} concludes the work and presents future work.

\section{Related Works}
\label{sect:relatedworks}
In recent years, plenty of research work has used SUMO. Some of them were trying to create an accurate scenario based on real traffic-data.

Vila Real Case Study \cite{VilaReal} presented a traffic scenario, where the traffic demand consisted of the evaluated census and survey data provided by governmental institutions, intending to estimate individuals' activities. An algorithm, named \textit{synthesizer} on this work, had the function to identify each region on census data and link it with a corresponding edge on SUMO. Unfortunately, this is a really small scenario restricted to the Portuguese census format, which cannot be applied to the data format provided by the City of Ingolstadt.

In 2013, a traffic signal control scenario within a realistic traffic simulation was presented \cite{Ottawa}. This scenario covers a 9 x 7 block section of Ottawa Downtown, and its main objective was to evaluate the effectiveness of an intelligent traffic control system based on a realistic scenario. The data used to generate the traffic was based on the turn values of each intersection. This study presented an approach and improvement results when an adaptive signal control system is implemented, increasing the simulation's average speed. Nevertheless, this scenario does neither cover an entire city behavior, nor the public transport network.

TAPAS Cologne \cite{TAPAS} is a scenario including the whole city of Cologne. It is based on real demographic data and inhabitant’s daily activity. Due to its size and complexity, it demands high computation time and still needs additional improvements in some features, e.g. junction corrections, correct lane numbers per street, adjust traffic lights position and insert public transport. At this scenario, traffic demand presents a realistic behavior, and at the same time road topology and public transport does not reflect the real-world equivalents, which makes it partly not realistic enough. 

Luxembourg scenario, LuST \cite{LUST}, is a high detailed scenario, which brings important features to implement VANET simulations, e.g. buildings. It was implemented and evaluated using demographic data and a measurement data-set, which consists of the average speed of some locations. This scenario includes all public transport networks and parking lots around the city, providing a more realistic traffic flow than for example the TAPAS Cologne scenario.

In 2017, the Monaco traffic scenario (MOST) \cite{MOST} was presented. This scenario showed land elevation characteristics, all public transport networks and also is multi-modal, considering not only vehicles but also bicycles and pedestrians. Unfortunately, this scenario covers only the morning peak hour and its traffic realism is not measured, because it was not compared to real traffic data.

All these previously presented traffic scenarios were developed by applying real information. Although, none of them include the traffic light system deployed by traffic authorities in a real traffic system, or have a robust validation method applying NRSME in different points of the city. These pitfalls will be pay attention in this scenario intending to improve traffic realism and mimic the Ingolstadt's traffic. Additionally, none of the above introduced scenarios cover the characteristics presented in Ingolstadt city, as:

\begin{itemize}
  \item An industrial city;
  \item One industry concentrating around 43\% of all work positions in one spot;
  \item Some companies working 24 hours a day in a 3 shift operation model;
  \item A high rate of vehicles per inhabitants~\cite{INGOLSTADT};
  \item A low unemployment rate~\cite{districts};
  \item A high income per inhabitant~\cite{districts};
  \item A higher rate regarding vehicle usage when compared to other cities~\cite{districts};
  \item Incoming traffic represents approximately 44\% of the total traffic \cite{verkehrspolitische};
\end{itemize}

\section{Ingolstadt Traffic Scenario (InTAS)}
\label{sect:InTAS}
\begin{center}
\begin{figure}[ht]
    \centering
    \includegraphics[width=0.6\textwidth]{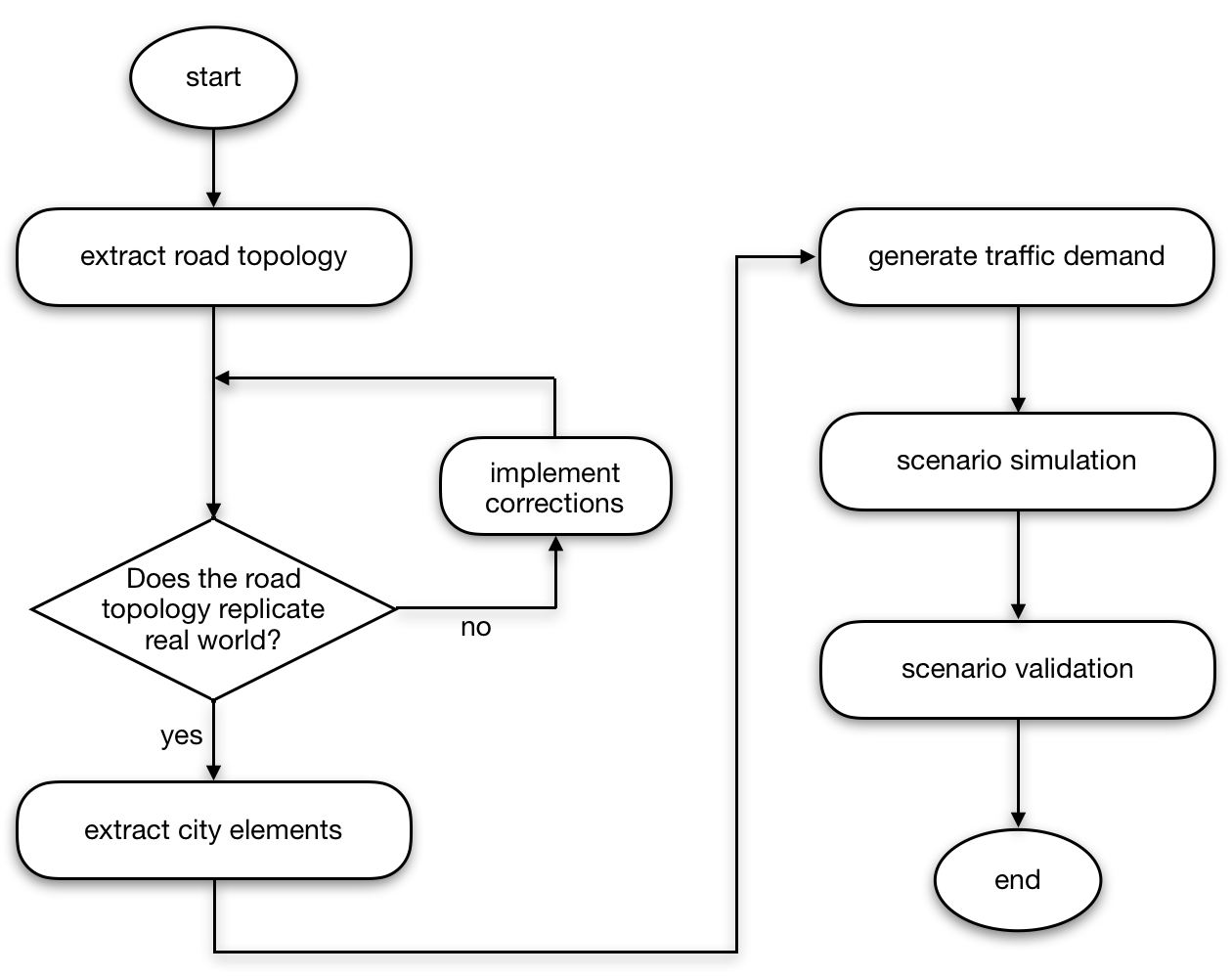}
    \caption{InTAS Flowchart}
    \label{fig:flowchart}
\end{figure}
\end{center}

The Ingolstadt Traffic Scenario (InTAS) development consisted of four different procedures: definition of network topology, traffic demand modeling, scenario simulation, and scenario evaluation. The objective of the first step was to elaborate a city map containing all important information, e.g. road topology, buildings, parking lots, bus stops and traffic light positions. The second step consisted of analyzing all available real-world data and identify a realistic traffic demand method that fits the data-set and creates a realistic traffic demand. The third step involved modifying the simulation parameters and preparation for the scenario evaluation. Figure~\ref{fig:flowchart} presents a flowchart of the developed process of InTAS.

\subsection{Map Creation}
\label{sect:mapcreation}

This stage is the foundation of the scenario and consisted in the scenario area delimitation. The chosen area represents 87\% of city’s work positions, approximately 79\% of the total inhabitants, and roughly 81\% of the registered cars in Ingolstadt. However, this selection excluded the traffic pattern of surrounding villages. This might not be an issue as some villages are over 12 $km$ away from the city center, and their inner traffic does not influence the main area of Ingolstadt. Thus, instead of modeling their internal traffic, this study took into consideration the traffic demand between the villages and Ingolstadt.

\begin{center}
\begin{figure}
    \centering
        \includegraphics[width=0.8\textwidth]{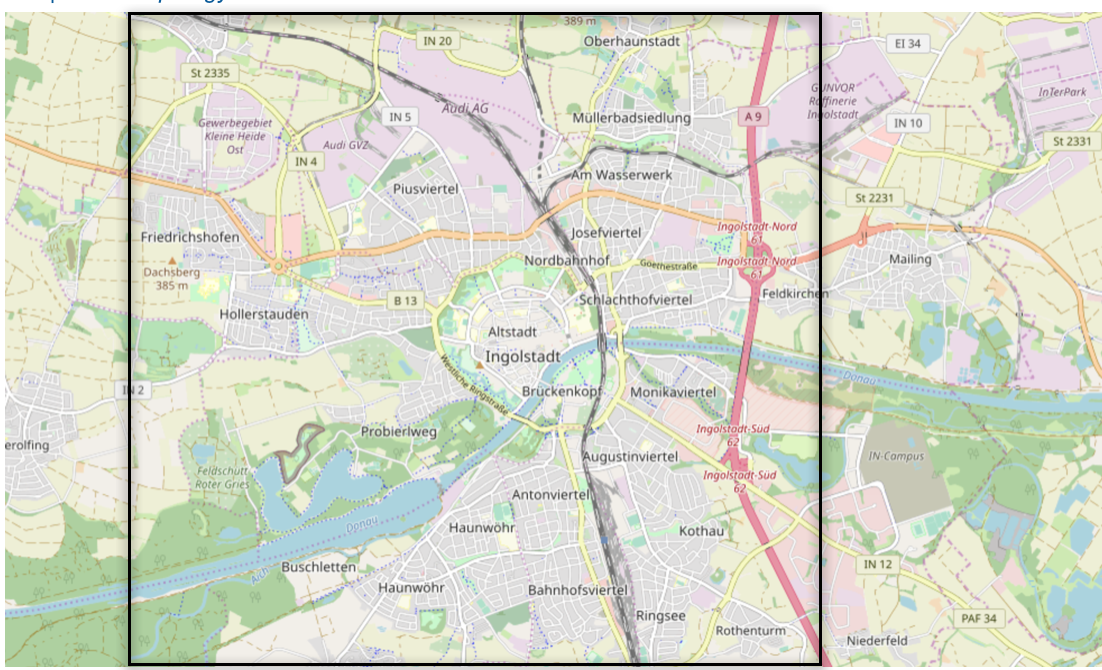}
    \caption{InTAS border of the selected area~\cite{Openstreetmap}}
    \label{fig:border}
\end{figure}
\end{center}

After selecting the interesting area, this region was extracted from OpenStreetMap (OSM) \cite{Openstreetmap} containing all information enclosed to this area. Figure~\ref{fig:border} shows the area selected for InTAS in OSM. 

\subsubsection{Road Network}
\label{sect:roadnetwork}

In SUMO, road networks are represented as \texttt{.xml} file grouped by the instances: edges and junctions. Edges represent road segments and junctions either correspond to intersections or dead-end streets \cite{SUMO}. As the OSM source file is presented with the \texttt{.osm} extension, it has to be converted into a SUMO readable format, \texttt{.xml}, applying the \texttt{netconvert} tool.

By examining the converted data, it was observed that a great number of streets was not representing the real-world, i.e. incorrect number of lanes, missing exclusive turn left lanes and exclusive bus lanes. This divergence might be caused by outdated information retrieved from OSM. Although information is frequently updated on OSM, as it is an open-source project working on the wiki-style process, some areas are not detailed enough, and, contain only the street segment but not the number of lanes or exclusive lanes. 

\begin{figure}
    \centering
    \begin{subfigure}[b]{0.3\textwidth}
        \centering
        \includegraphics[width=\textwidth]{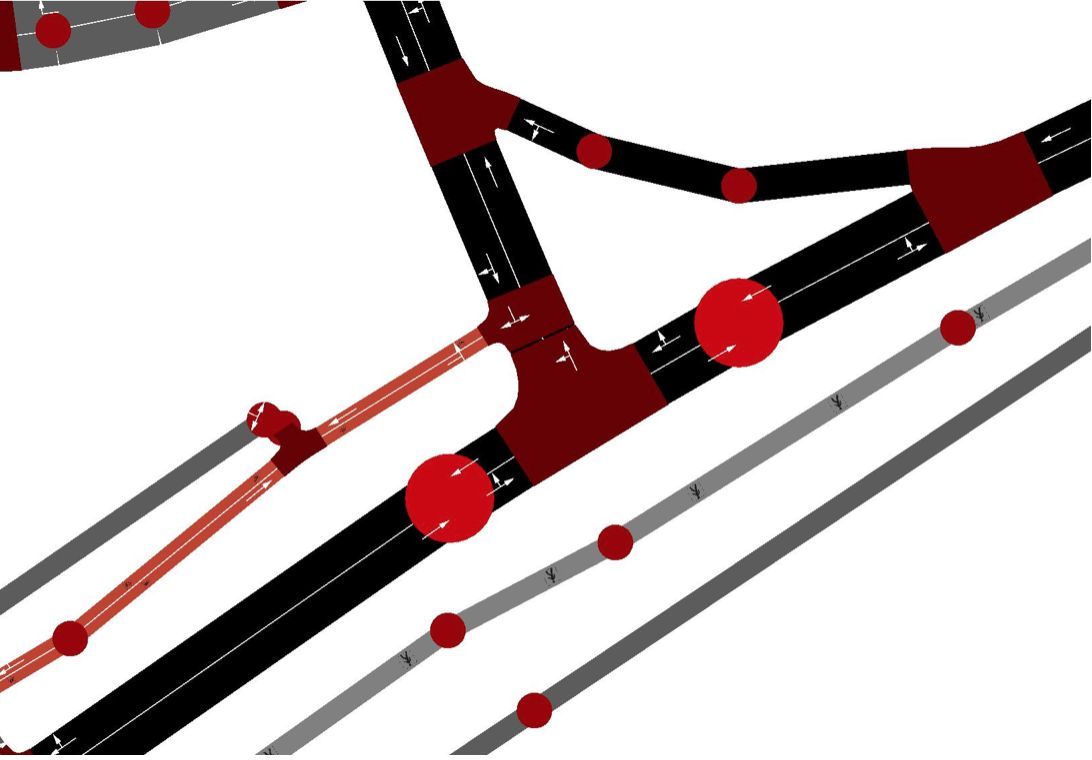}
        \caption{Conversion Result}
        \label{fig:firsimport}
    \end{subfigure}
    \begin{subfigure}[b]{0.3\textwidth}
        \centering
        \includegraphics[width=\textwidth]{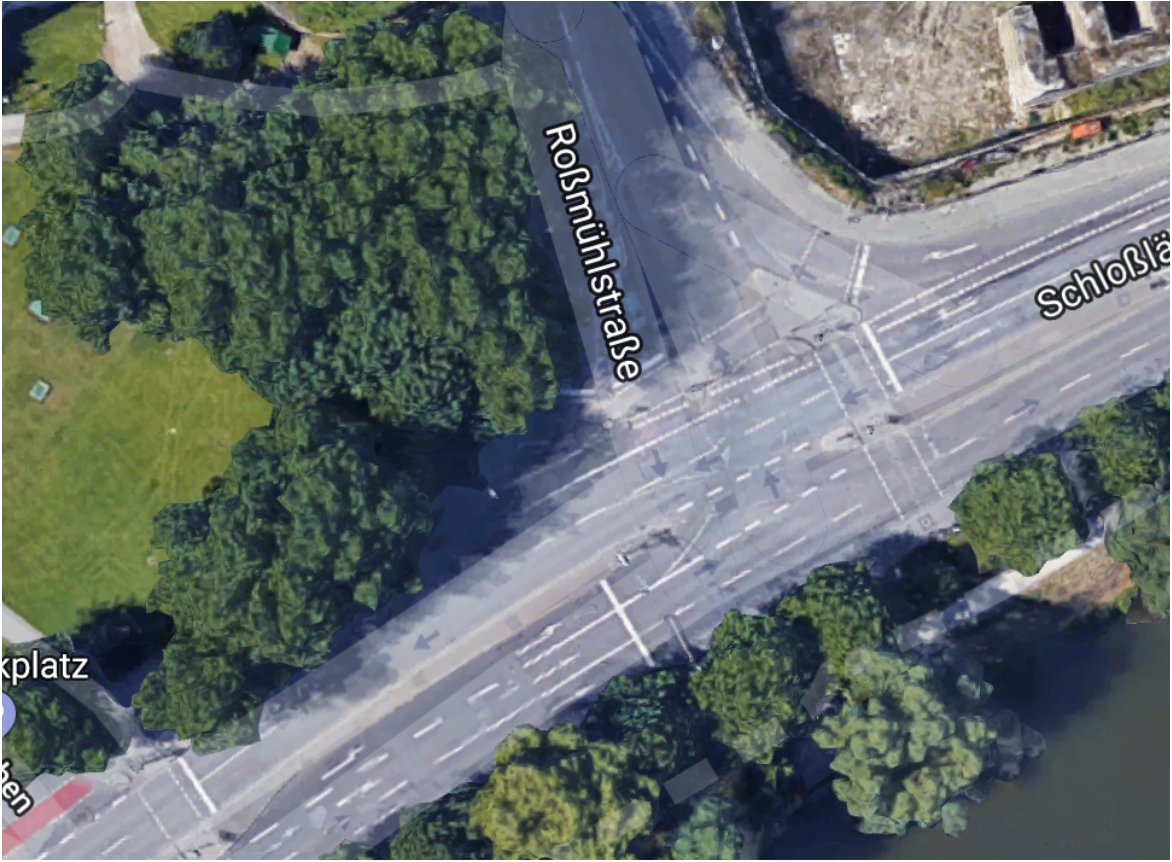}
        \caption{In Google Maps~\cite{google-maps}}
        \label{fig:googlemaps}
    \end{subfigure}
    \begin{subfigure}[b]{0.3\textwidth}
        \centering
        \includegraphics[width=\textwidth]{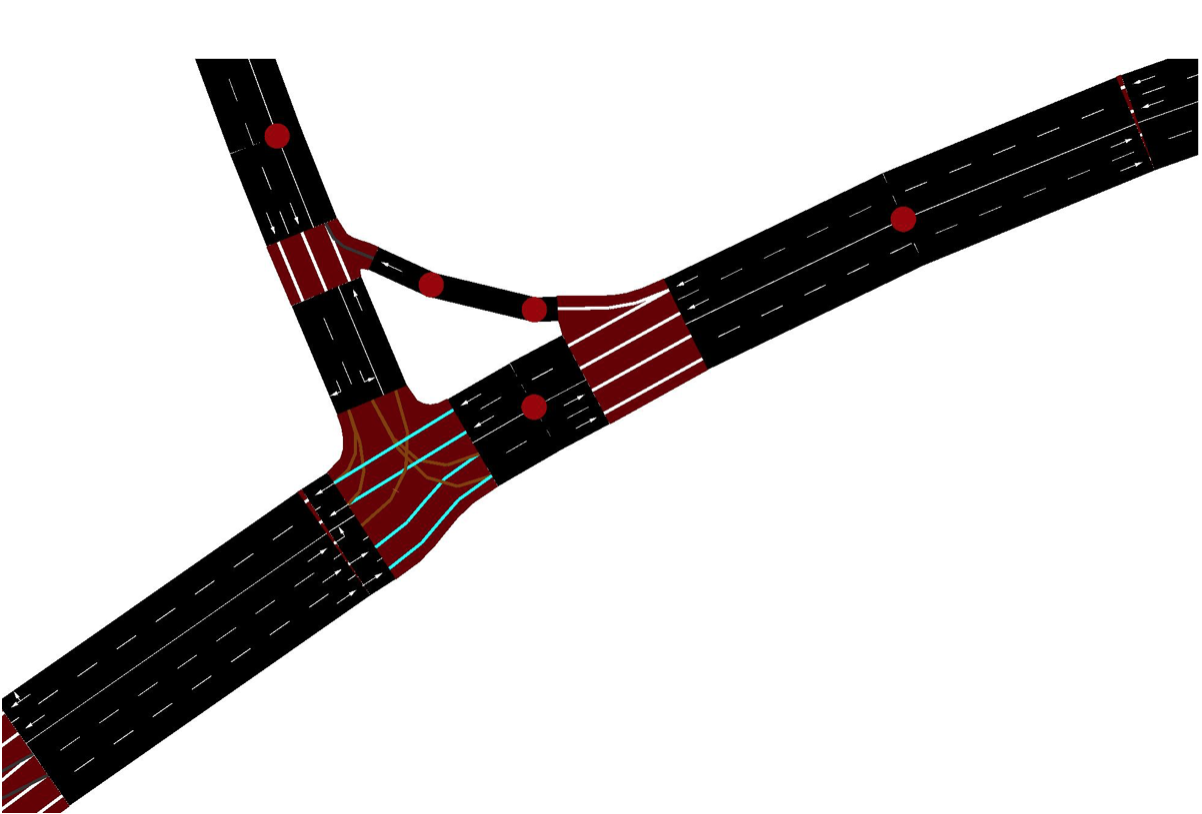}
        \caption{Manually Corrected}
        \label{fig:correction}
    \end{subfigure}
    \caption{Road adjustments in their three steps}
    \label{fig:adjust}
\end{figure}

Intending to develop a reliable map, which accurately represents Ingolstadt, it was necessary to implement a method to correct all the issues. Thereby, a thorough process to compare each of the 7,966 edges and all of the 3,341 nodes with the satellite image, on-line accessible, on Google Street Maps \cite{google-maps} was undertaken. This correction were applied with the \texttt{netedit} tool, where the entire map was manually inspected and validated. During the correction, all junctions were checked to reinforce lane connections. Moreover, all bicycle lanes, sidewalks, and private streets, as commercial shopping facilities, residential and industrial condominiums, were removed. After the cleaning process, the total road length of the InTAS scenario is 717.13 $km$. In Figure~\ref{fig:adjust}, partial results from this action are shown. Figure ~\ref{fig:firsimport} is the first conversion result, Figure ~\ref{fig:googlemaps} is the same road in Google Maps and Figure ~\ref{fig:correction} is the final result after editing the area by hand.

As the radio propagation model might differ from tunnels, under-buildings, and under-bridge passages, e.g. in the free space propagation~\cite{radioPropagationModel}, this paper utilizes road categories to inform the communication simulator. Thus, it provides the categories \texttt{tunnel}, \texttt{under.building}, and \texttt{under.bridge}.

Traffic light systems also play an important roll in the city traffic behavior \cite{trafficlight}. Due to their importance, they are considered for the scenario. As \texttt{.osm} files contain the traffic light positions, \texttt{netconvert} assembled it together with the road network's \texttt{.xml} file, assigning a hypothetical phase length to all traffic lights (TL). At this step, the objective was to check if all real TL were represented on the map. To confirm all TL positions, an up-to-date document provided by the City of Ingolstadt, containing all locations from TL managed by them, was used. Furthermore, all pedestrian-only TLs were removed from the map, keeping just those that control crossings with a minimum of two streets junction, resulting in a total of 98 traffic lights across the map.

\begin{figure}
    \centering
        \includegraphics[width=0.8\textwidth]{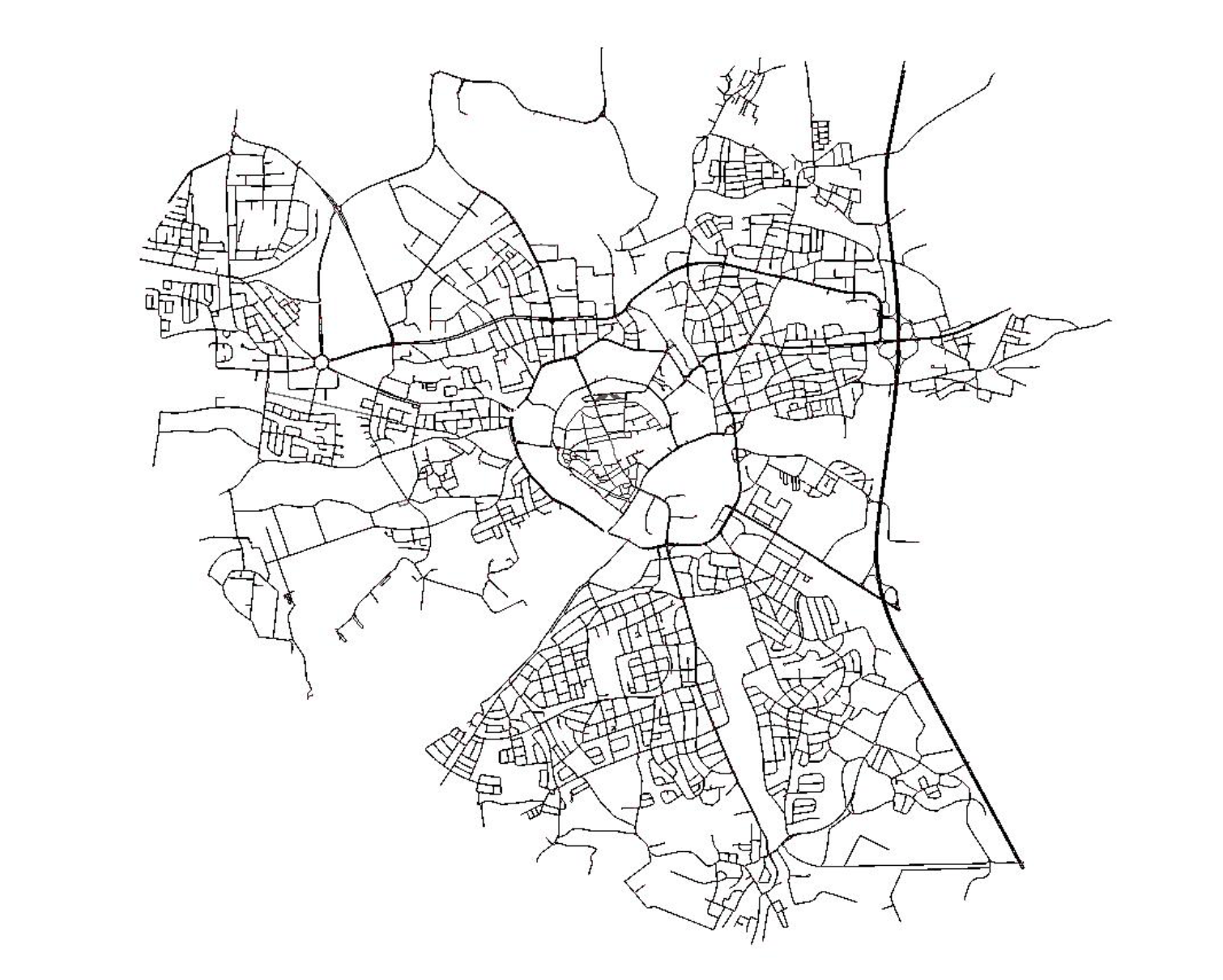}
    \caption{InTAS Road Topology}
    \label{fig:topology} 
\end{figure}

The final InTAS' road network, after implementing all the corrections and adjustments, is presented in Figure~\ref{fig:topology}. Table~\ref{tab:roadnetwork} shows the information concerning to road network developed at this stage of the development.

\begin{table}[tb]\small
\begin{center}
  \begin{tabular}{lcc}
   \hline
    Parameter               & & Value\\
   \hline
    Total Area              & & 51.54 $km^{2}$  \\ 
    Road Length             & & 717.23 $km$ \\
    Nodes                   & & 3,342  \\
    Edges                   & & 7,968 \\
    Traffic Lights          & & 98 \\
   \hline
  \end{tabular}
\end{center}
\normalsize
\caption{Network numbers
  \label{tab:roadnetwork}}
\end{table}

\subsubsection{City Elements – Parking, Traffic Lights, Buildings and Bus Stops}
\label{sect:elements}

Due to the great number of variables that directly impact traffic behavior it is extremely complex to model the city traffic. This section presents additional considerations for the scenario: parking areas, traffic lights, buildings, and bus stops.

\paragraph{Parking Areas.}
In the \texttt{.osm} data, 59 parking areas are represented, but not all of them were used in this work. InTAS has focused on public parking areas and companies' parking. The City of Ingolstadt manages a total of 13 parking areas with 5,568 parking lots. These parking areas were tracked during an average usage at business days between Tuesday to Thursday, from September to December 2019. This measurements were used, resulting in daily average usage of 4,247 public parking lots in the city. This average value was inserted in the simulation, to know the average number of vehicles to park there. The parking areas currently closed for construction work were considered to be always full, which partly reflects their usage before getting closed.

Moreover, additional five parking areas were considered in the simulation. Three serving AUDI AG, one Klinikum Ingolstadt and one Technische Hochschule Ingolstadt (THI). For THI, only the number of employees and not the students was considered. The total workers from those three employers represent 54.52\% of the total employees in the scenario, where AUDI AG is the largest company in Ingolstadt, employing 44,526 workers, Klinikum Ingolstadt employs 3,630 and THI employs 650 workers. These areas were identified in the map and assigned with the number of employees for each one.

\paragraph{Traffic Lights.}
When importing TLs using \texttt{netconvert}, a generic TL program is automatically generated and assign to each traffic light, defining the traffic light cycle time, each phase duration, states, and the traffic light logic type.

TL state is the definition linked for each lane under a Traffic Light System (TLS) operation, i.e. assigning if the light is green, red or amber for this lane. TLS cycle is the total time necessary for a program to run all phases. Phase duration is the time a state will be activated. The parameter TL logic type may assume \textit{static} or \textit{actuated} values. A \textit{static} parameter represents a TL with consistent behavior, never changing phase durations.TL logic type was set as \textit{actuated} traffic control, extending a phase once continuous traffic is detected~\cite{SUMO}.

As TLS might be one of the highest influencing factor to traffic behavior~\cite{trafficlightimpact}, this work seeks to provide a realistic TLS to its junctions, therewith, to near simulation traffic to real traffic. In the scenario 20 TLS have been simulated based on the real program deployed on the real traffic lights. The others TLs were implemented with the automatically generated program applying the \textit{actuated} traffic logic.

\paragraph{Bus Stops.}
InTAS also reproduces the public transport system, considering all bus lines inside the simulation area. Yet, to provide more realistic representation, all bus stops were imported from the \texttt{.osm} file. Those stops were compared with the online available information from local bus service company - Ingolst\"adter Verkehrsgesellschaft (INVG)~\cite{INVG}. A total of 404 bus stops were inserted to the scenario.

\paragraph{Buildings.} 
InTAS was developed to be a complete simulation environment, where traffic behavior as well as VANET studies can be deployed. To encompass a higher simulation potential, this scenario implemented all buildings represented by the \texttt{.osm} file. Due to their influence in the performance for an inter-vehicular wireless communication environment, when operating in the standardized frequency, they are very important \cite{Buildings}. As aforementioned, the \texttt{.osm} data is a resourceful database, and all information related to buildings was converted applying the \texttt{polyconvert} tool. A total number of 21,756 buildings were incorporated into InTAS.

\begin{figure}
    \centering
        \includegraphics[width=0.8\textwidth]{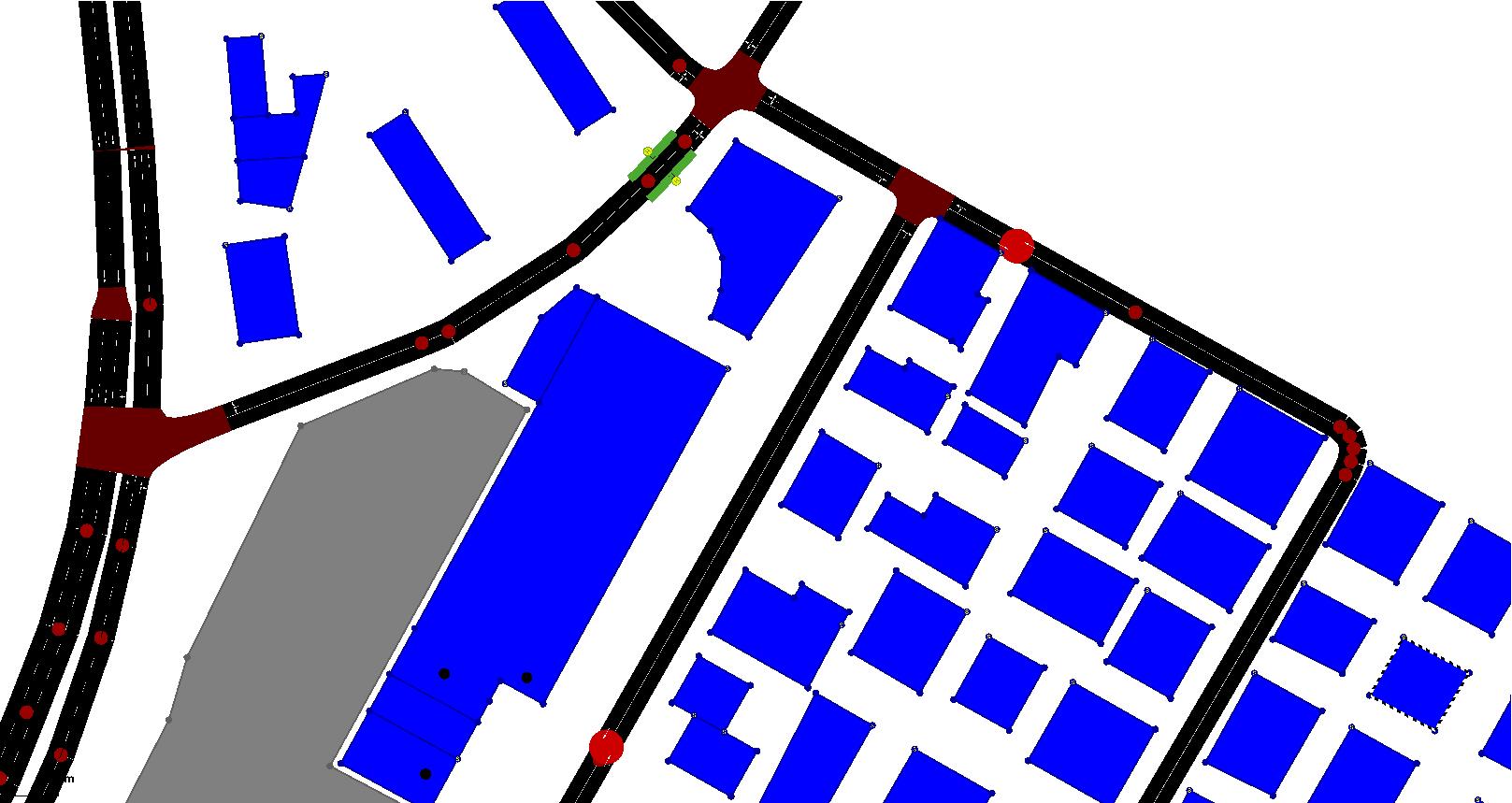}
    \caption{Extract of InTAS with its city elements}
    \label{fig:elements} 
\end{figure}

\paragraph{}
Figure~\ref{fig:elements} presents an extracted part of the Ingolstadt map, where the road network is shown together with all city elements. In blue color, buildings are represented with their dimensions and shapes. The gray color indicates parking lots. Bus stops are marked in green aside the roads.  

\subsection{Real Traffic Database}
\label{sect:realtrafficData}

In interaction with the Ingolstadt Verkehrsmanagement und Geoinformation Office, which is a branch of the City of Ingolstadt, an SFTP server with information from 24 traffic measurement points have been structured. For each point, information between September 3\textsuperscript{rd} 2019 and December 15\textsuperscript{th} of the same year, has been taken into consideration. The available data describes the number of vehicles which daily drove through these areas over 24 hours of the day, grouping the total number of vehicles in a 15 minutes time window.

For each measurement point, values for Tuesdays, Wednesdays, and Thursdays were selected. These values compute the heaviest traffic days according to the Ingolstadt Verkehrsmanagement und Geoinformation Office. Among the selected days, holidays and days before the holiday were excluded, because the traffic may change its characteristic at these days. Data from days that faced any issue have been also removed. In the end, each crossing remained with data from 27 days{\footnote{average number of days for measurement point.}}. Thereafter, the average value for each detector in each junction has been calculated. Computing the average value avoids choosing a day with unusual behavior, e.g. working sites and snowy days.

A data-set, which has been split into two sub-sets, was utilized to provide information for modeling and validation. One for modeling the traffic demand and evaluate simulation parameters, which consisted in data from October 2019. The other sub-set has been used in the validation step, which took into consideration traffic values from November 2019.

\subsection{Traffic Demand}
\label{sect:trafficdemand}

The second step to elaborate the traffic scenario, was the insertion of moving vehicles on the previously developed map. This objective, known as traffic demand, defines the number of vehicles in the simulation, their origin and destination edges, departure time, and the path they will drive through to reach their destination, i.e. which are the consecutive edges between the start and end edges. At this point, SUMO distinguishes traffic demand in trips and routes. Trips represent a general view from traffic demand and it is a model only containing edge of origin, the edge of destination and departure time, i.e. the local this traffic is originated and the local the destination is. In contrast, a route is an expanded view, representing, besides the origin and destination edges, all edges that the vehicle transits through, i.e. a path is assigned to the origin and destination and involves all edges during its flow.~\cite{SUMO}

\subsubsection{Traffic Demand Modeling}
\label{sect:trafficmodeling}

\begin{figure}
    \centering
        \includegraphics[width=0.8\textwidth]{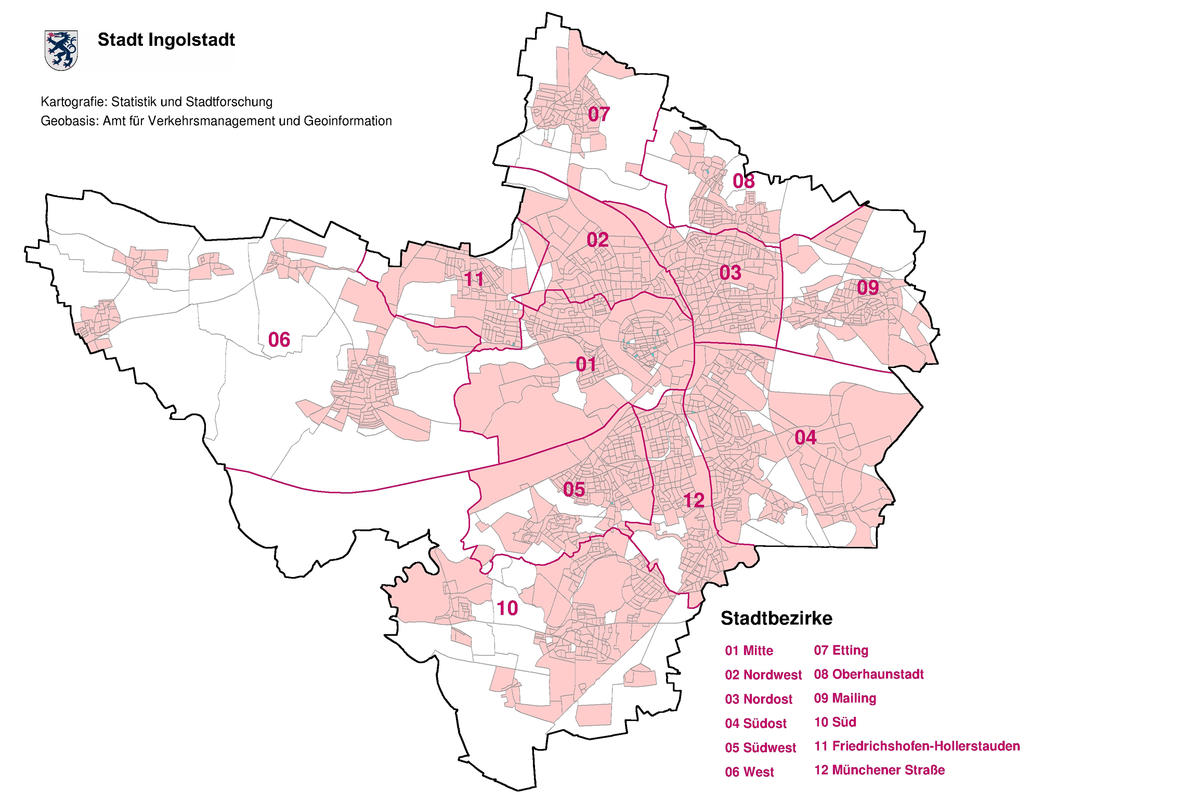}
    \caption{Districts of Ingolstadt City (Source:~\cite{subareas})}
    \label{fig:district} 
\end{figure}

Ingolstadt is divided into 12 districts, where each of these areas are subdivided, creating a total of 62 sub-districts, as shown in Figure~\ref{fig:district}, where districts are numbered from 1 up to 12 and sub-districts are delimited by the gray line inside the district. For each of these sub-districts, based on data from the City of Ingolstadt, it is known: number of inhabitants, households, living workers, unemployed, and number of registered vehicles. These numbers are at a high level of detail, providing a reliable database to model Ingolstadt's traffic. 

According to the available data, real traffic data presented in Section~\ref{sect:realtrafficData} and online demographic data from the City of Ingolstadt, it has been decided to model InTAS applying the \texttt{activitygen}~\footnote{\url{https://sumo.dlr.de/userdoc/Demand/Activity-based_Demand_Generation.html}} method. 

In the developed statistic file for \texttt{activitygen}, mostly attributes related to general information, parameters, population's age brackets, and schools have been calculated based on online demographic data provided by the City of Ingolstadt. However, parameters related to traffic information, as \texttt{incomingTraffic}, \texttt{outgoingTraffic}, and \texttt{cityGates} have been modeled based on data-set with real traffic numbers from October 2019. 

Amongst the scenario, 38 out of 62 sub-areas have been considered laying within the InTAS borders. The inhabitants for these selected areas were divided into 13 age groups, ranging from 0 to above 85 years. Thereafter, the number of social numbers presented on each sub-area was computed, intending to define the number of workers that live in each region and the total number of working positions per area. Table~\ref{tab:demographicnumbers} presents the difference between the total Ingolstadt demographic numbers~\cite{districts} and the demographic numbers applied in InTAS. In Table~\ref{tab:demographicnumbers} attribute workers refers to the number of workers living inside the area. Based on these numbers, a difference of approximately 21\% related to a number of inhabitants and workers between entire Ingolstadt and InTAS is observable. The difference detected for the number of vehicles and householders is smaller and is nearly at 18\% for both. However, a tinier difference is noticed for work positions and unemployed, showing 13\% and 7\% respectively.

The difference between city's number and InTAS might influence the traffic behavior and the number of vehicles driving through the map. To solve this issue, the traffic demand generated outside the InTAS' border, concerning to Ingolstadt city, was considered as incoming traffic to the scenario. A deficit between working positions and workers inside the scenario was also observed. Therefore, the same solution, considering this as incoming traffic, was applied.

\begin{table}[ht]\small
\begin{center}
  \begin{tabular}{lcc}
   \hline
   Attribute                            & Ingolstadt City    & InTAS \\
   \hline
    Inhabitants                         & 138,180            &  109,090 \\ 
    Workers                             & 61,670             &  49,020 \\
    Work positions                      & 102,925            &  89,515 \\
    Unemployed                          & 1,219              &  1,138\\
    Vehicles                            & 97,950             &  80,337 \\
    House holders                       & 69,379             &  57,118 \\

   \hline
  \end{tabular}
\end{center}
\normalsize
\caption{Comparison of demographic numbers between Ingolstadt and InTAS
  \label{tab:demographicnumbers}}
\end{table}

Not only incoming traffic is relevant for an inner traffic city, but also outgoing traffic. Intending to model this phenomenon, it was important to define the gates, through which the traffic comes and leaves the scenario area. Based on available traffic information, a total number of 22 points, where the traffic can income and outgo from the scenario, have been defined~\cite{gevas}. Moreover, it was primordial to assign the number of vehicles incoming and outgoing through each one of this gates. With given data from the City of Ingolstadt, each gate was defined with its own traffic flow, representing a total number of 55,374 incoming people and 14,879 outgoing people to their work. Furthermore, the car preference rate of 0.589 has been defined, representing the probability an inhabitant to uses the vehicle instead of using other transportation means~\cite{SUMO}. Table~\ref{tab:trafficnumbers} summarizes vehicle traffic numbers and also presents the car rate, which describes the rate of adults that own a vehicle inside the scenario area~\cite{SUMO}.

\begin{table}[tb]\small
\begin{center}
  \begin{tabular}{lrlr}
     \hline
   Attribute                            & Value\\
    \hline
    Car rate                  &  0.9363 \\ 
    Incoming traffic          &  55,374 \\ 
    Outgoing traffic          &  14,879 \\
    Car preference rate       &  0.5890 \\
   \hline
  \end{tabular}
\end{center}
\normalsize
\caption{Vehicle traffic numbers for InTAS~\cite{verkehrsentwicklungsplan}
  \label{tab:trafficnumbers}}
\end{table}

Companies' opening and closing hours are also relevant for modelling the traffic. This information was assigned considering the proportion of workers that have to start and finish their jobs at that time. For companies that work 24 hours uninterruptedly the start and end of shift time were considered as opening and closing times. The data provided by the city is based on measurements of a normal business day. Thus, Tuesdays, Wednesdays or Thursdays were selected. According to the traffic management office, these days are the busiest traffic days and were taken into consideration for traffic improvements.

Children also play an important role in traffic demand. Although most of them do not go to work, a multitude of them is driven to kindergarten or school by their parents. To represent this behavior in the Ingolstadt scenario, each school was defined, containing their exact position on the map, the age range it covers, capacity and class hours. Thus, this step includes children from the kindergarten age to high school age. Moreover, to include students from both universities, Technische Hochschule Ingolstadt (THI) and Katholische Universität Eichstätt-Ingolstadt (KU), a similar implementation was designed, but at this point, parents do not drop them off. Instead, they drive their own vehicles. Likewise the workers, but with the universities as the final destination. InTAS considered 17 kindergartens, 36 schools, and 2 universities.

All structured demographic data were used to define the trips, i.e. identify where people live, work and study. The number of trips reached by this study was 333,741, considering the entire traffic for 24 hours of all vehicles. Although, this information was not sufficient to describe the path each driver will take to achieve his destination. For this reason, the~\texttt{duarouter} tool was used, which is the application to assign an entire path between origin and destination points, computing the routes for each vehicle.

\subsubsection{Simulation of Public Transport System}
\label{sect:publictransport}

Bus lines are an important factor in real-world city traffic, as they influence the traffic behavior of all participants. Especially on one lane driving roads, when they have to stop at a bus stop. Therefore, this work also considers the bus lines as they drive through Ingolstadt. Busses' behavior can be represented in SUMO, where it is possible to set drivers behavior, bus routes, and simulate busses stopping at their stations. However, simulating vehicles passing stopped busses is not possible. Thus, the average time a bus spend in each stop has been set to 10 $seconds$~\cite{LUST}, which approaches to the main objective.  

In Section~\ref{sect:elements} it was presented how bus stops were extracted of a~\texttt{.osm} file and how they were converted into a \texttt{.xml} file. To represent a realistic public transport system, all bus routes for this scenario were considered. Intending to typify bus routes, it was resorted again to INVG public data. Thus, it was possible to feed the real busses route information to the simulation. In total, there are 56 bus lines, where 28 are regular lines, 15 are night lines, 7 are shift lines --- running only at specific period --- and 6 are lines that attend surrounding cities, but have their departure or arrival in Ingolstadt. The bus lines cover a total of 880.6 km of road and compute 1,620 bus trips during the 24 hours simulation. Parameters of public transportation are shown in Table~\ref{tab:bus}.

\begin{table}[tb]\renewcommand{\tabcolsep}{10pt}
\begin{center}
  \begin{tabular}{lr}
    \hline
    Attribute                            & Value \\
   \hline
    Number of lines         &  56 \\
    Total bus stops         &  404 \\
    Number of busses trips  &  1,620 \\
    Bus routes length       &  880.6 km \\
   \hline
  \end{tabular}
\end{center}
\normalsize
\caption{Public transport numbers~\cite{INVG}
  \label{tab:bus}}
\end{table}

\subsection{Traffic Flow Optimization}
\label{sect:trafficOptimization}

Implementing \texttt{duarouter}, the assignment performance computes the shortest path through the network using the Djikstra~\cite{dijkstra} route-planning algorithm. At this point, \texttt{duarouter} defines the shortest route for each vehicle, considering that they are alone on the road network. Due to this, after loading all the vehicles in the simulation, it will lead to traffic jams.

Seeking to mitigate this issue, an equilibrium state might be reached. For this reason, the Gawron's \cite{Gawron} method to optimize the traffic flow has been implemented. This method calculates the user equilibrium for each vehicle and implement a route optimization. SUMO provides the tool \texttt{duaIterate}, which iteratively tries to find the user equilibrium, i.e. to find a route for each vehicle without reducing the travel cost~\cite{SUMODUA}. As the number of iterations to reach the equilibrium might vary, the parameters average speed, time lost, and travel time were analyzed for InTAS.

\begin{figure}
    \begin{subfigure}{0.325\textwidth}
        \includegraphics[width=\textwidth]{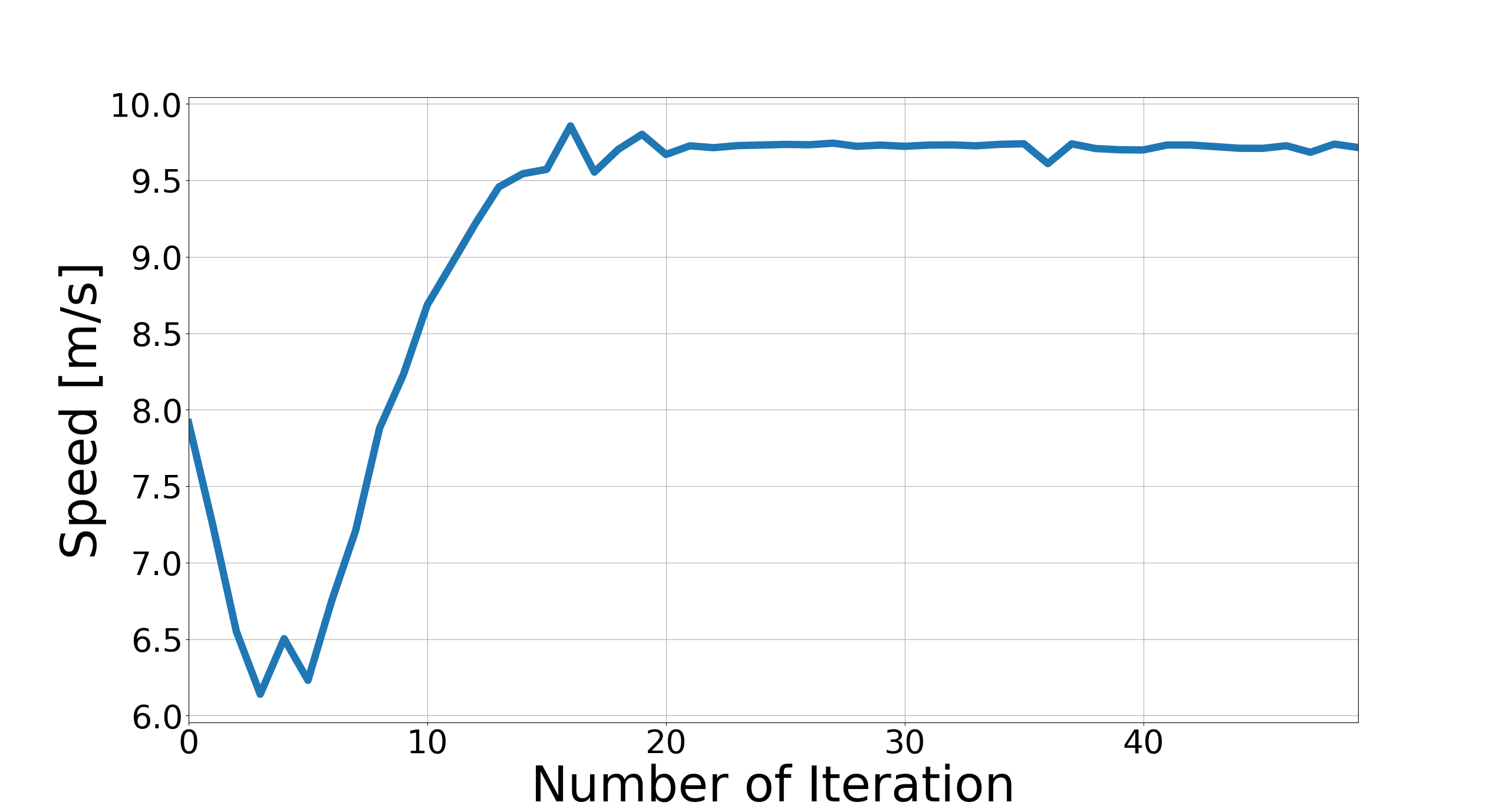}
        \caption{Average Speed}
        \label{fig:avgSpeed}
\end{subfigure}
\begin{subfigure}{0.325\textwidth}
    \centering
        \includegraphics[width=\textwidth]{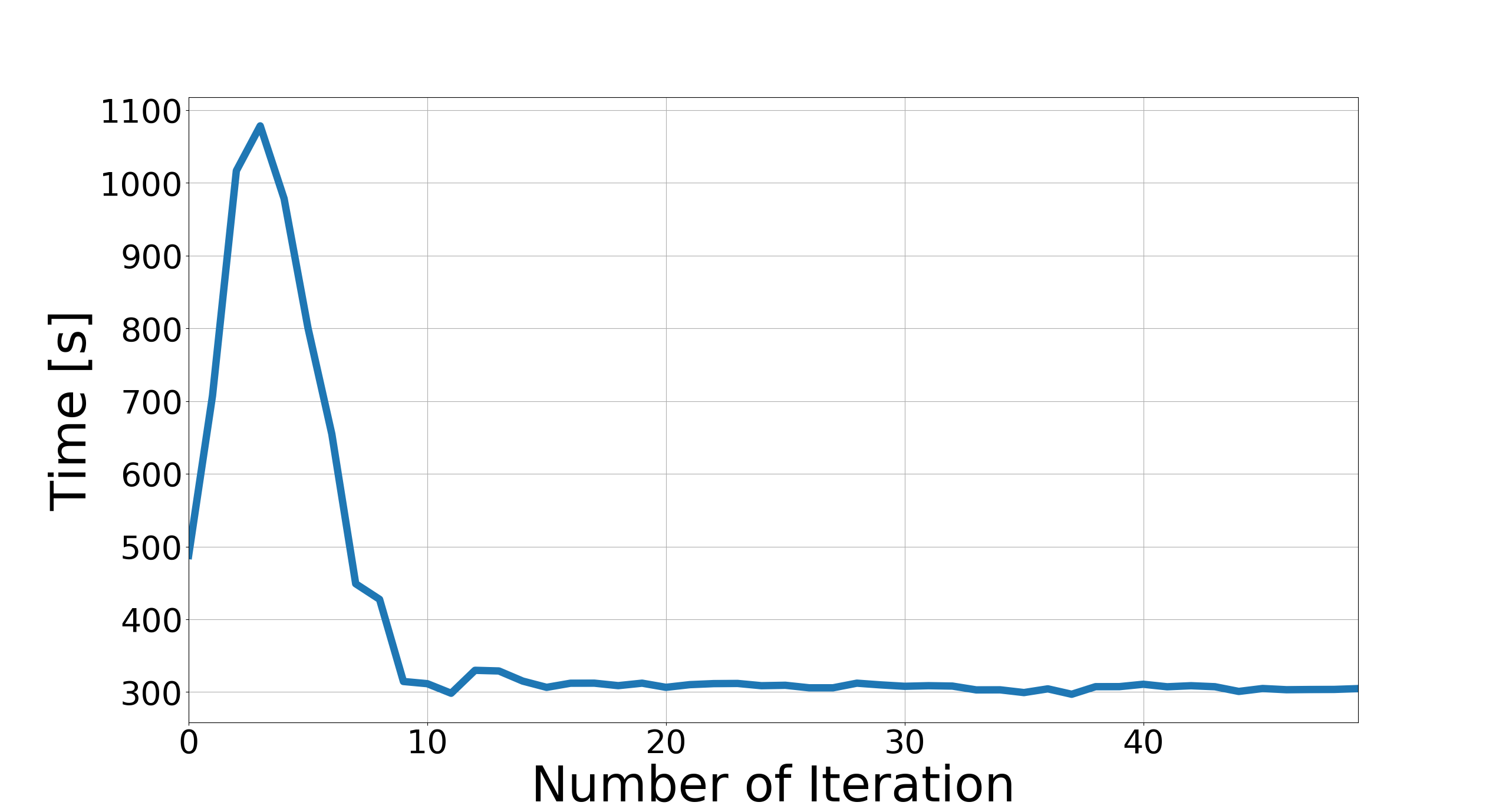}
        \caption{Time Lost}
        \label{fig:timeLoss}
\end{subfigure}
\begin{subfigure}{0.325\textwidth}
    \centering
        \includegraphics[width=\textwidth]{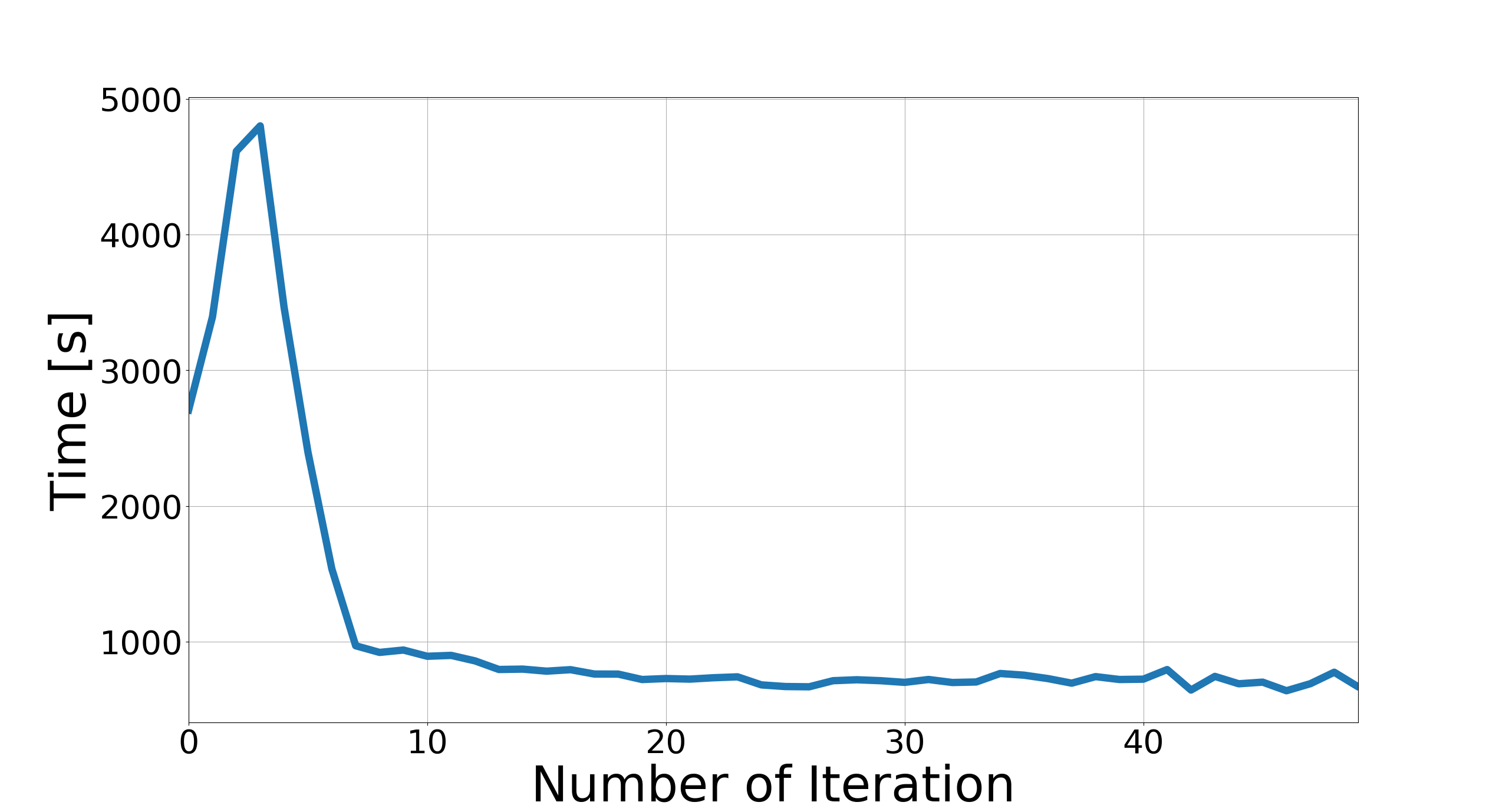}
        \caption{Average Travel Time}
        \label{fig:avgTravelTime}
\end{subfigure}
\caption{InTAS Iterations}
\label{fig:stabilization}
\end{figure}

Time loss, average speed, and average travel time are parameters which tend to converge to a stability value when the equilibrium is reached. Figure~\ref{fig:stabilization} shows the result obtained after 50 iterations applying \texttt{duaIterate} for the parameters. It can be observed that the equilibrium has been achieved for all parameters after 25 iterations. Figure~\ref{fig:avgSpeed} describes the average speed presented in the scenario for each iteration, demonstrating an unexpected behavior in the initial iterations. This behavior reduced InTAS' average speed to a minimum value of 6.15~$m/s$ on the 3\textsuperscript{rd} iteration, and only from the 6\textsuperscript{th} iteration, the average speed grows up to the optimum value. Between 8\textsuperscript{th} and 22\textsuperscript{nd} iterations an oscillatory behavior is observed. Only on the iteration 25, the equilibrium has been reached with an average speed of 9.73~$m/s$.

Figure~\ref{fig:timeLoss} exhibits the parameter time loss for each iteration. This parameter is provided by the SUMO simulation and calculates, for each vehicle, the difference between the actual trip duration and theoretical trip duration~\cite{SUMOTimeLoss}. On the 9\textsuperscript{th} iteration is noticed that the time loss reached 316.6~\textit{seconds}, and decreases slower, when compared to the initial iterations, up to 308.5~\textit{seconds} in the iteration 25. Figure~\ref{fig:avgTravelTime} shows the average travel time for each travel in each iteration, converging to 931.5~\textit{seconds} on the 9\textsuperscript{th} iteration. These three figures also present that in the first iterations the traffic scenario was denser and with a low mobility pattern. However, after iterations of \texttt{duaIterate}, the equilibrium state has been reached. Therefore, since all parameters converged in the 25\textsuperscript{th} iteration, it has been assumed to proceed with the further development of InTAS.

\subsection{InTAS Simulation}
\label{sect:simulation}

Simulation is the phase where scenario map file, moving vehicles represented in route file, and additional files as bus stations, buildings and detectors files are gathered. These files are regulated by parameters defined at this step.

Simulation parameter \texttt{begin} and \texttt{end} time were set to cover 24 hours of a day, with a \texttt{step-length} of 0.1 seconds. Usually, it may happen that a vehicle blocks an intersection, which leads to a huge traffic jam, causing an unrealistic pattern in the simulation. Avoiding such behavior, the parameter \texttt{ignore-junction-blocker} allows vehicles to ignore a junction after a specific time and continue their travel from there. A value of 15 $seconds$ was set for this feature, intending to minimize the impacts. Another setting feature is \texttt{time-to-teleport}, which defines the maximum vehicle’s waiting time in seconds on a traffic jam before it is teleported to a further position of its own route, intending to reduce impact created for huge traffic jams. For this, the parameter was set to the default value presented by SUMO, which is 300 $seconds$. In the simulation settings, it is also possible to define the routing algorithm and vehicle following model. As vehicle following model, the \textit{Krauss} model~\cite{krauss} was defined, which models the reaction times and human behavior during the drive, introducing a stochastic component, e.g. the driver’s behavior when changing lanes. Table~\ref{tab:parameters} summarizes the simulation parameters used by InTAS.

\begin{table}[ht]\renewcommand{\tabcolsep}{10pt}
\begin{center}
  \begin{tabular}{lr}
   \hline
    Parameter                   & Value\\
   \hline
    begin                               & 0 seconds  \\
    end                                 & 86,400 seconds  \\
    step-length                         & 0.1 second \\
    ignore-junction-blocker             & 15 seconds \\
    time-to-teleport                    & 300 seconds \\
    default.carfollowmodel              & Krauss \\
    routing-algorithm                   & Dijkstra \\  
    device.rerouting.probability        & 0.82 \\    
    device.rerouting.period             & 300 seconds \\
   \hline
  \end{tabular}
\end{center}
\normalsize
\caption{Simulation Parameters
  \label{tab:parameters}}
\end{table}

Another feature presented in the simulation phase in SUMO is the parameter named \texttt{device.rerouting.probability}, which allows vehicles to change their routes during the simulation. In real-world traffic, some drivers may change their path, due to the knowledge they have about the city traffic. To address this behavior, this parameter was applied to this scenario. To calculate the best rate for \texttt{device.rerouting.probability}, an algorithm has been developed to search the optimum value.

\subsubsection{Defining InTAS Best Rerouting Probability}
\label{sect:reroutingProbability}

SUMO has a variety of simulation parameters, which influences the traffic behavior. As each city has its characteristics, each parameter may change from city to city. Among these parameters, there is the \texttt{device.rerouting.probability} representing the probability of a vehicle to have a rerouting device. Vehicles equipped with this device may compute a new route as soon they come across an unexpected situation, like a traffic jam that can increase the time cost to reach the destination.

Changes to the extreme on this parameter will lead the traffic behavior to two distinguishes performances. Setting it to null represents that all vehicles will drive through the same roads - edges. The performance noticed is that the traffic jams will increase until SUMO crashes due to hardware limitations to deal with the high number of vehicles running in the simulation. Another issue faced at this point is that with more vehicles in the simulation they could not reach their destination, and that is the reason why the number of vehicles keeps growing. On the other hand, setting \texttt{device.rerouting.probability} to 1.00 will provide a large traffic capillarity. That will force vehicles to use roads that are not often used and will reduce the number of vehicles using the main roads, inducing an unrealistic pattern.

Intending to find the best value for \texttt{device.rerouting.probability} that fits for InTAs, an algorithm to iterate this parameter from 0.00 to 1.00 with an iteration-step of 0.01 has been developed. The algorithm has compared the total number of cars driving through each measurement point presented in the data-set with its respective value in the simulation applying Normalized Root Mean Square Error (NRMSE),

\begin{equation}
    NRMSE = \frac{\sqrt{\frac{\sum_{n=1}^{N}(x_{r,n} - x_{s,n})^2}{N}}}{\Bar{x}}
    \label{eq:RMSE1}
\end{equation}where $x_{r,n}$ represents point-values for the first sample, $n$ and $x_{s,n}$ represents the values for the second sample at the same time-window. $N$ represents the total number of samples. $\Bar{x}$ represents the mean value of the measured data. Based on the equation~\ref{eq:RMSE1}, it is possible to infer that lower the NRMSE value, lower the error between the series.

\begin{figure}
    \centering
    \includegraphics[width=0.8\textwidth]{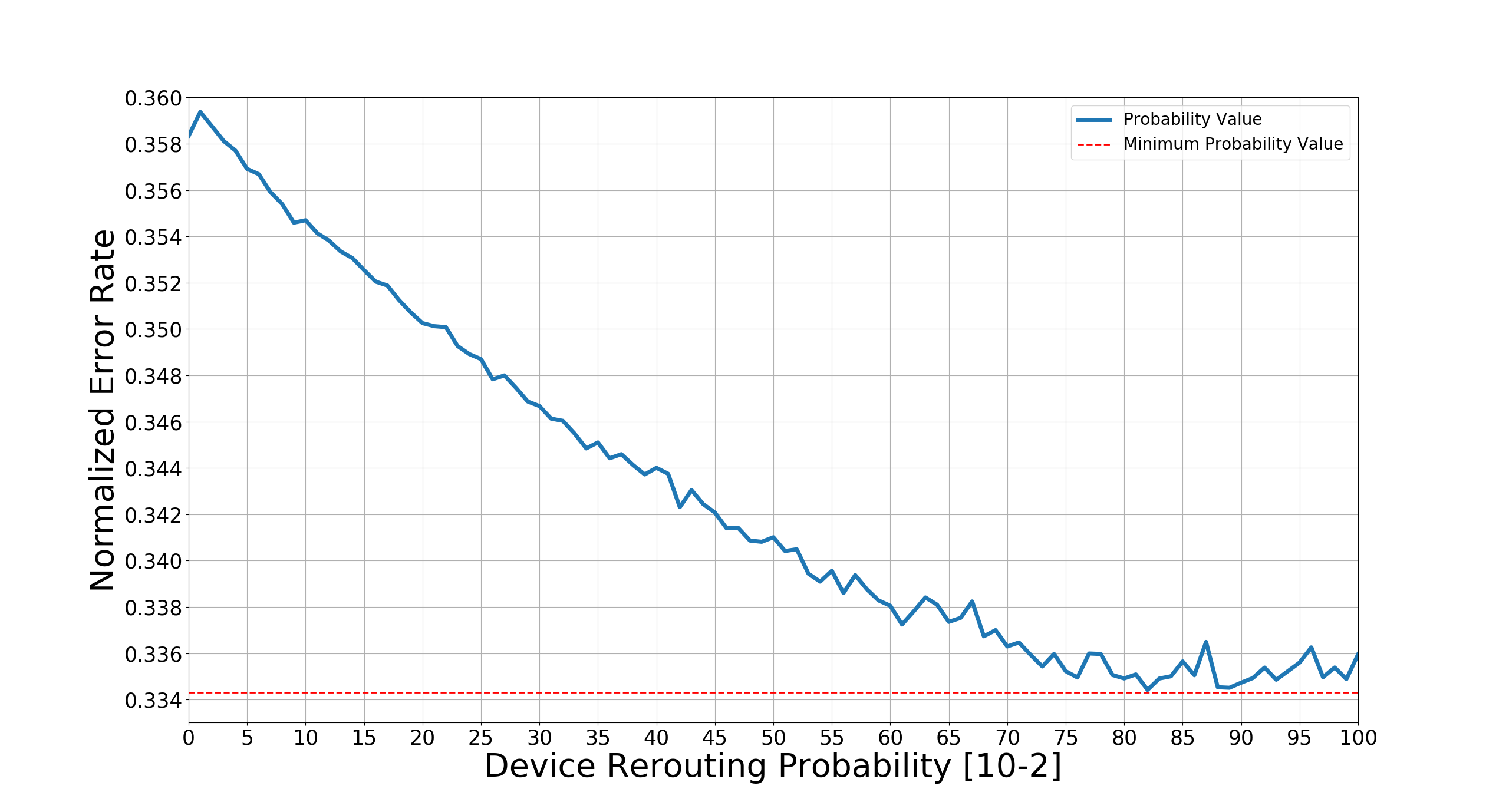}
    \caption{Device Rerouting Probability Evaluation}
    \label{fig:rateStudy} 
\end{figure}

Figure~\ref{fig:rateStudy} shows the behavior of error rates obtained by the simulations, where the error is higher for low values, decreases over the iteration until reaches the lower error value, and raises again until the end of the iterations. The best value reached for \texttt{device.rerouting.probability}, which presents the lower error rate, is the simulation with 0.82 of probability with an NRMSE, i.e. an error rate of 0.3343. This value was considered for validation analysis presented in Section~\ref{sect:evaluation} and as the final value for this parameter in InTAS.

\subsubsection{InTAS Behavior}
\label{sect:trace}

\begin{figure}
    \centering
        \includegraphics[width=0.8\textwidth]{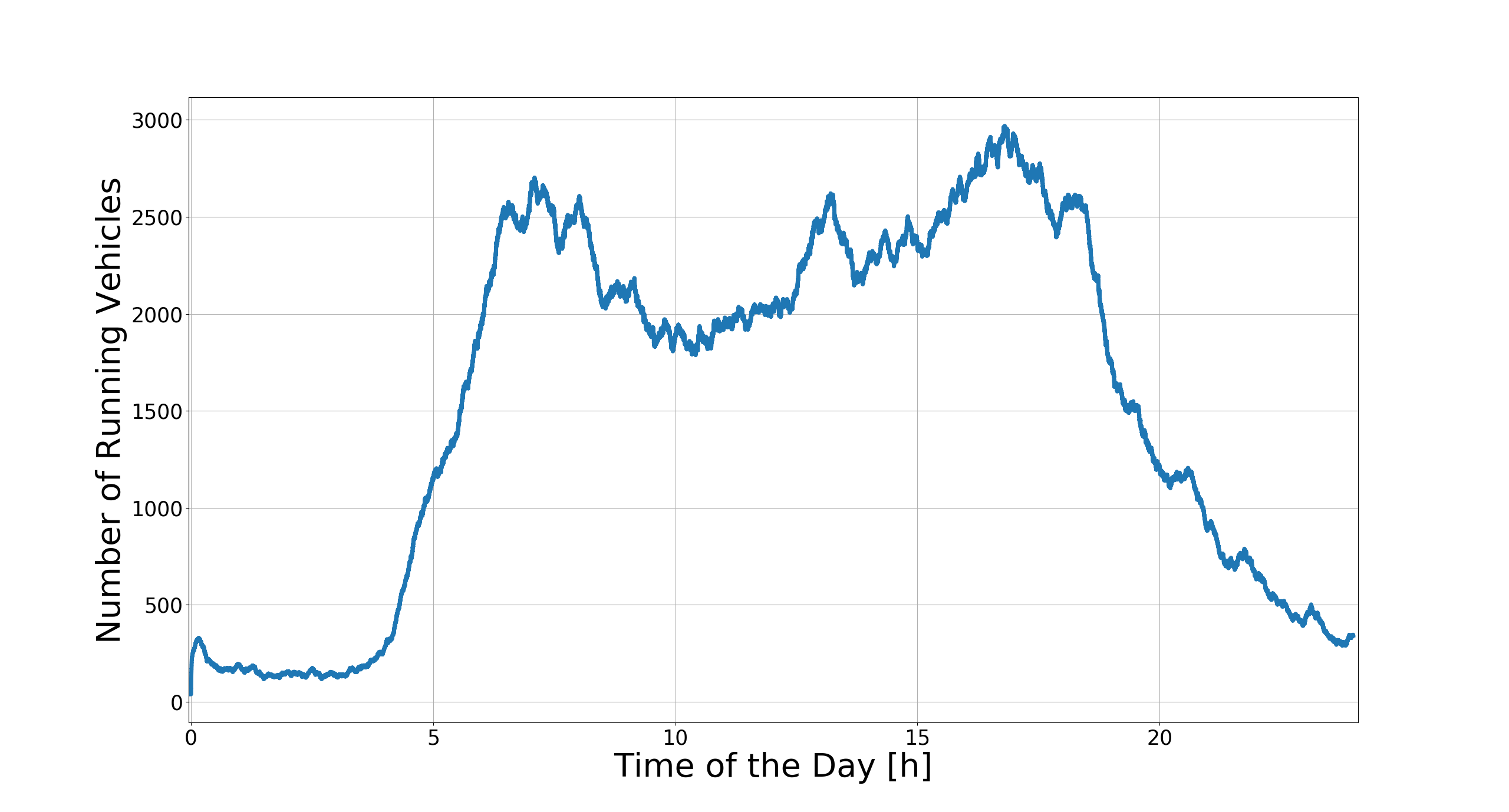}
        \caption{Running Vehicles for InTAS}    
        \label{fig:behavior} 
\end{figure}

Figure~\ref{fig:behavior} depicts InTAS's traffic behavior. Where the first peak starts right before 4:00, representing the flow for the beginning of the first shift gathered with the end of the night shift. Traffic is still increasing until around 6:25 and remains in the morning peak up to 8:12 with approximately 2,500 vehicles on the streets. The morning peak time is a bit before when most offices start their activities and is also correlated with students. After reaching the morning peak, the traffic behavior starts decreasing until 10:25, when the lowest number of vehicles driving around the scenario after the morning peak is observed. This valley computes 1,795 vehicles on the simulation. After this time the number of vehicles grows until 12:24, where a peak in this growth behavior is notice. This noon peak lasts from 12:24 up to 13:42, which can represent people going for lunch and the finish of morning classes. After lunchtime peak the number of vehicles still growing, representing the end of the first shift and the beginning of the second shift. The phenomenon slightly increases the number of running vehicles until the afternoon peak around 16:47, which presents the highest number of running vehicles for InTAS, with 2,965 in total. Afterward, the number of vehicles decreases until the end of the day, with a slight peak from 20:14 to 21:05, representing the end of the second shift and the beginning of the third shift.

\section{InTAS Validation}
\label{sect:evaluation}
\begin{figure}
    \centering
        \includegraphics[width=0.8\textwidth]{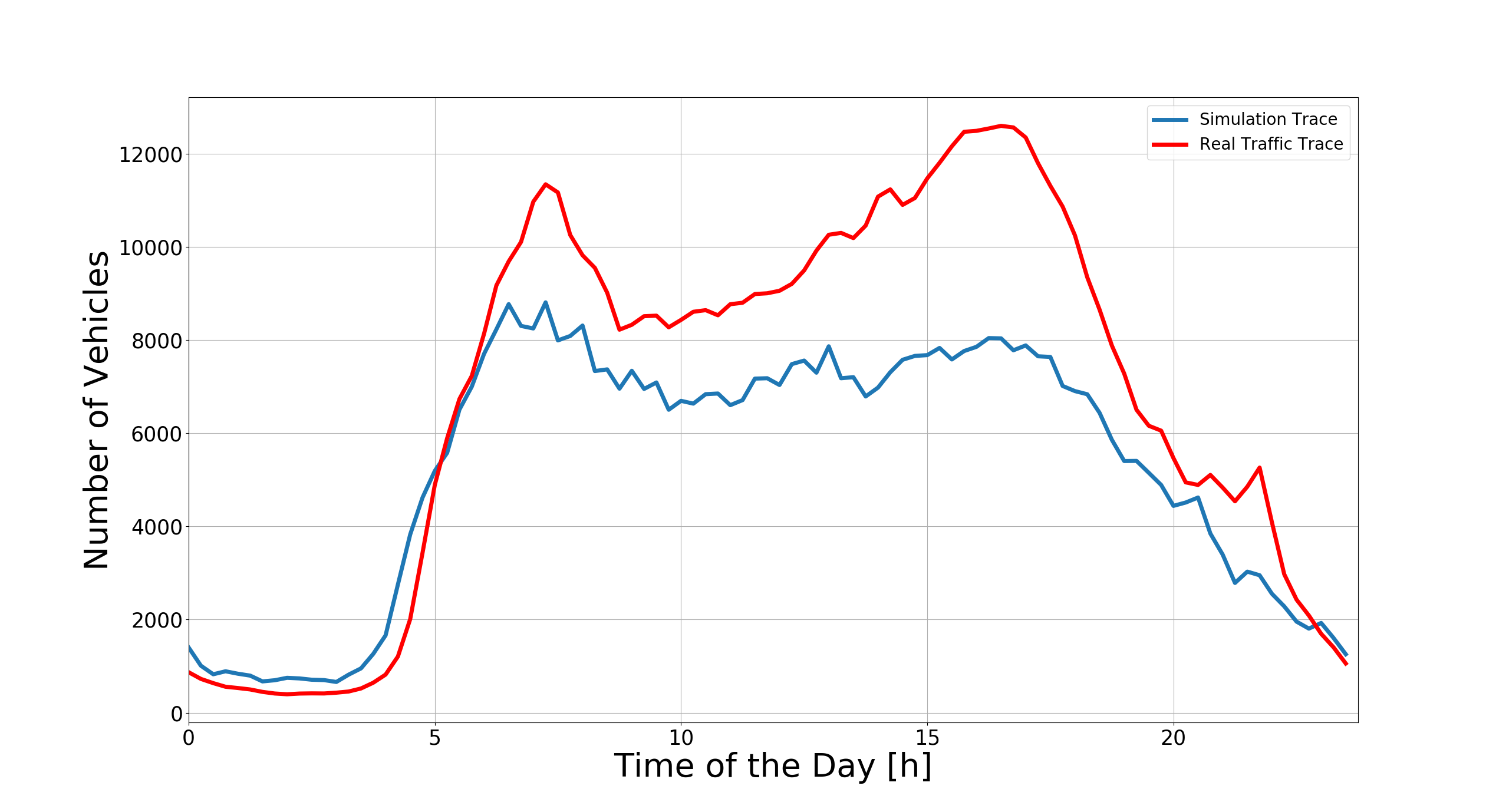}
        \caption{Traces Comparison}
        \label{fig:junction_compare} 
\end{figure}

The validation of the InTAS scenario was done based on the data-set presented in Section~\ref{sect:realtrafficData}, considering the detector values from November 2019. The comparison between all detectors is depicted in Figure~\ref{fig:junction_compare}, where the blue line represents the trace resulting from the simulation, and the red line shows the values from the data-set. The mismatch calculated applying NRMSE brings a rate of 0.33 for the scenario. In the Simulation Trace (ST), a large number of vehicles are observed from 0:00 until 5:06 when compared with the Real Trace (RT). The RT trace exceeds ST from 5:06 to 22:52. After this time, a lower mismatch between the traces is observed. Around 17:00 the highest peak is detected, and thereafter the number of vehicles in both traces reduces. Starting at 22:52 both traces have a similar behavior until the end of the day.

\begin{figure}
    \begin{subfigure}{0.49\textwidth}
        \includegraphics[width=\textwidth]{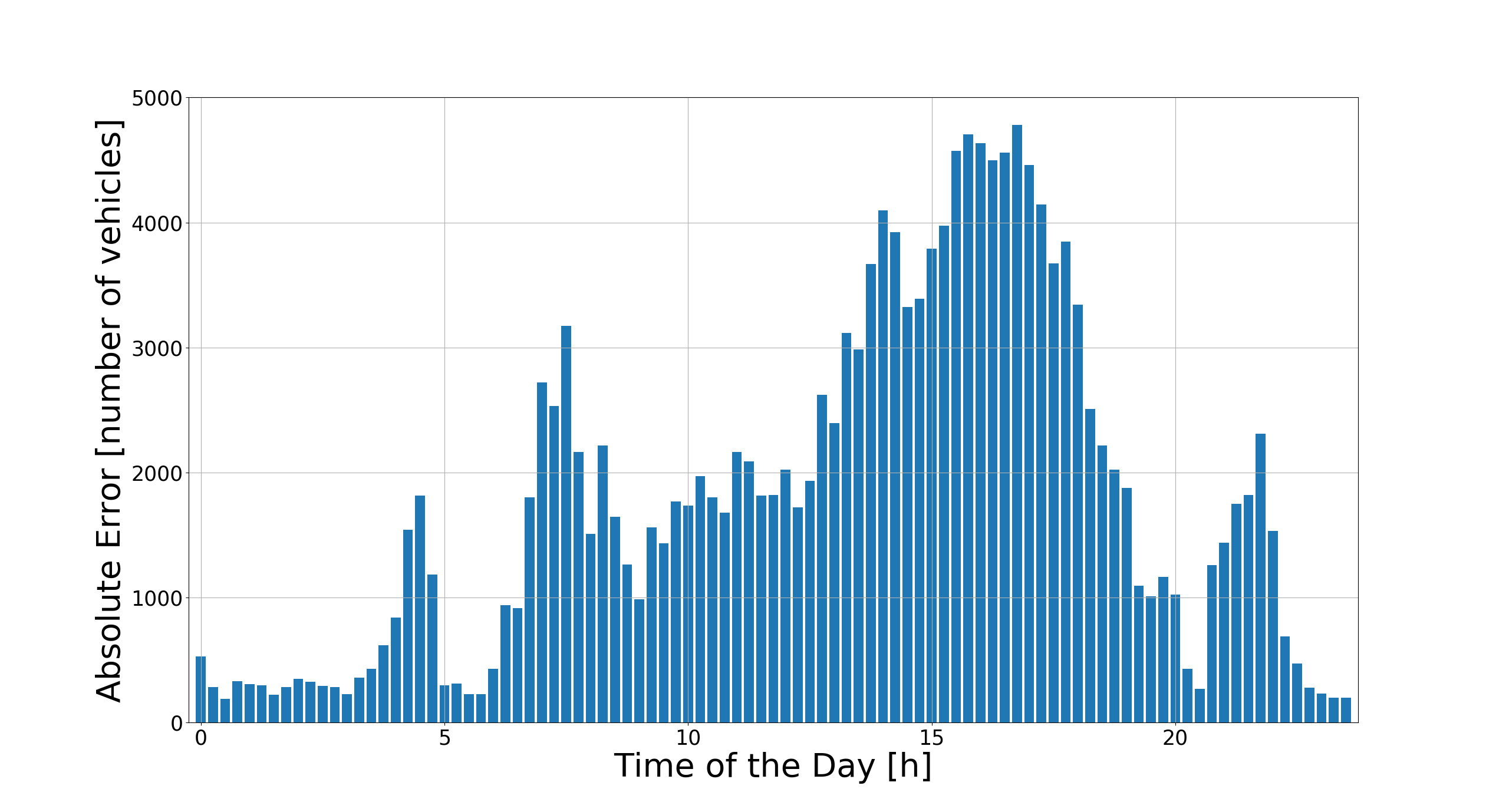}
        \caption{Absolute Error}
        \label{fig:rse} 
    \end{subfigure}
    \begin{subfigure}{0.49\textwidth}
        \includegraphics[width=\textwidth]{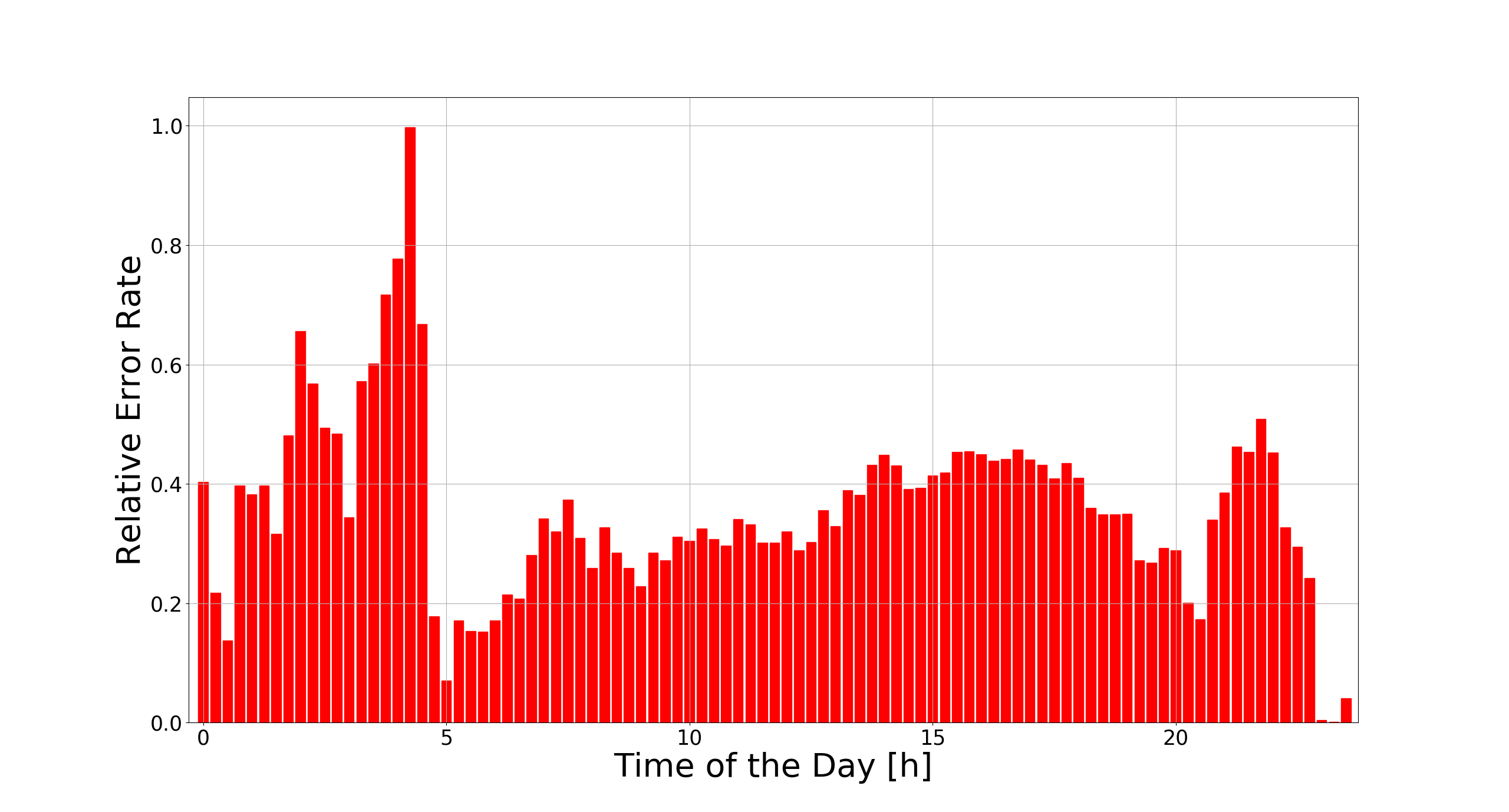}
        \caption{NRMSE}
        \label{fig:rer} 
    \end{subfigure}
    \caption{Error Analysis}
    \label{fig:errorAnalysis}
\end{figure}

Intending to enrich the analysis and to understand the traffic behavior, an absolute error has been calculated. This error considers the absolute difference for each of the time samples, comparing simulation and real values. Figure ~\ref{fig:rse} shows the absolute error behavior during the day, and it is possible to imply that the highest error is computed between 15:30 and 17:00 in the afternoon. At this time, a higher number of vehicles is computed, and the simulation did not follow the same behavior presented on the data-set. The lowest errors occur at the time when traces cross to each other, and during the period from 0:00 to 3:30, between 8:45 and 9:30, and after 22:15.

Although Figure~\ref{fig:rse} shows that the highest absolute error occurs between 15:30 and 17:00 in the afternoon, it is necessary to measure the influence caused by the absolute error values. Therefore, an analysis comparing the NRMSE for each time window is depicted in Figure~\ref{fig:rer}. As can be seen, the highest absolute error is between 15:30 and 17:00, with an NRMSE of about 0.45, which is 27.3\% higher than the scenario's NRMSE. The error presented between 1:45 and 4:30 in the morning has a larger impact, even that it has a smaller absolute error, when compared with other periods of the day. 

\subsection{Crossing Evaluation}
\label{sect:crossingEvaluation}

A total number of 24 measurement points were compared, and no pattern to describe the error has been observed. Due to this, the intersections with the highest error and the one with the lowest error were deeper analyzed.

\begin{figure}
\begin{subfigure}{0.48\textwidth}
    \centering
        \includegraphics[width=\textwidth]{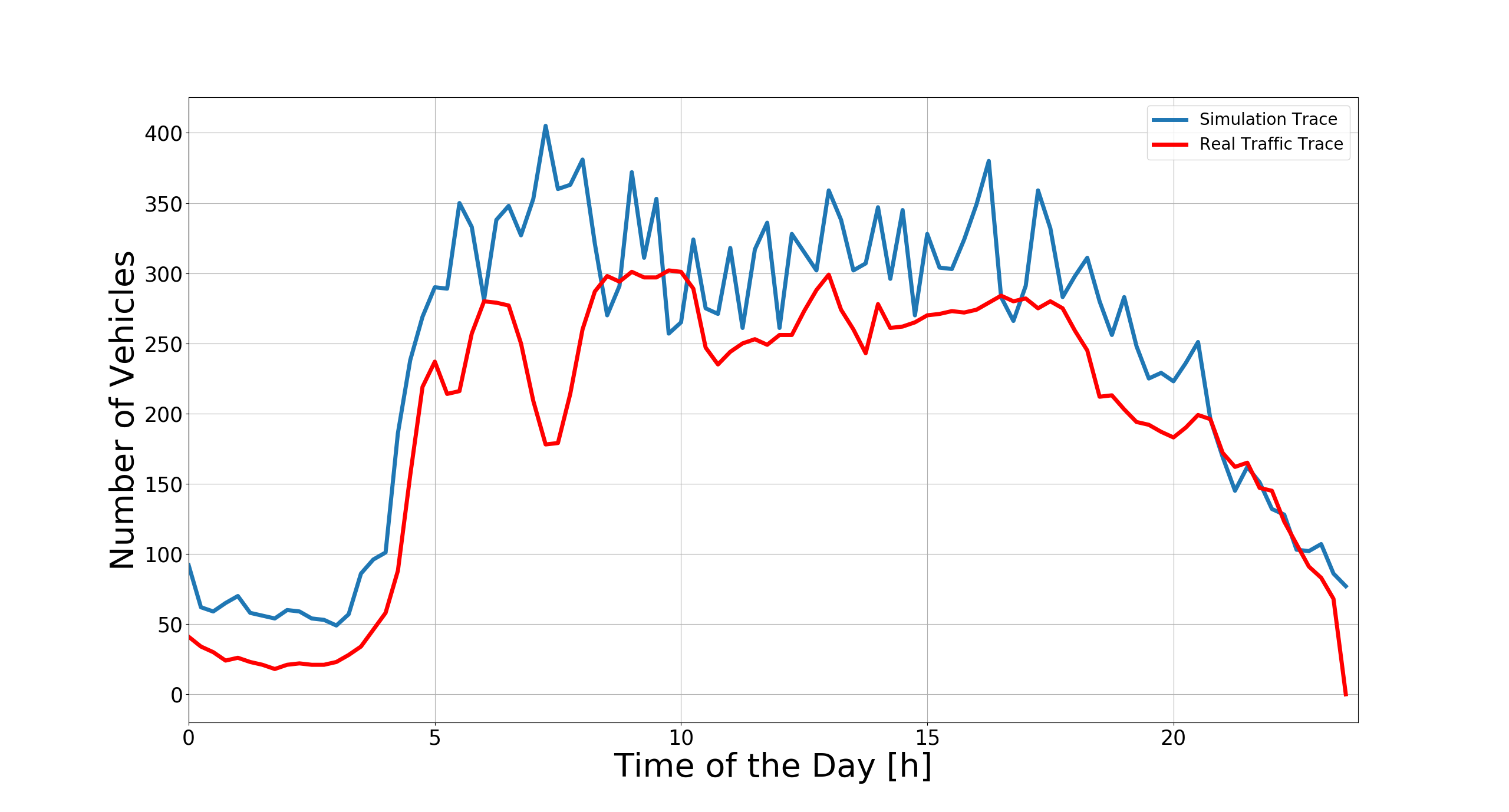}
        \caption{Best Case}
        \label{fig:4210} 
\end{subfigure}
\begin{subfigure}{0.48\textwidth}
    \centering
        \includegraphics[width=\textwidth]{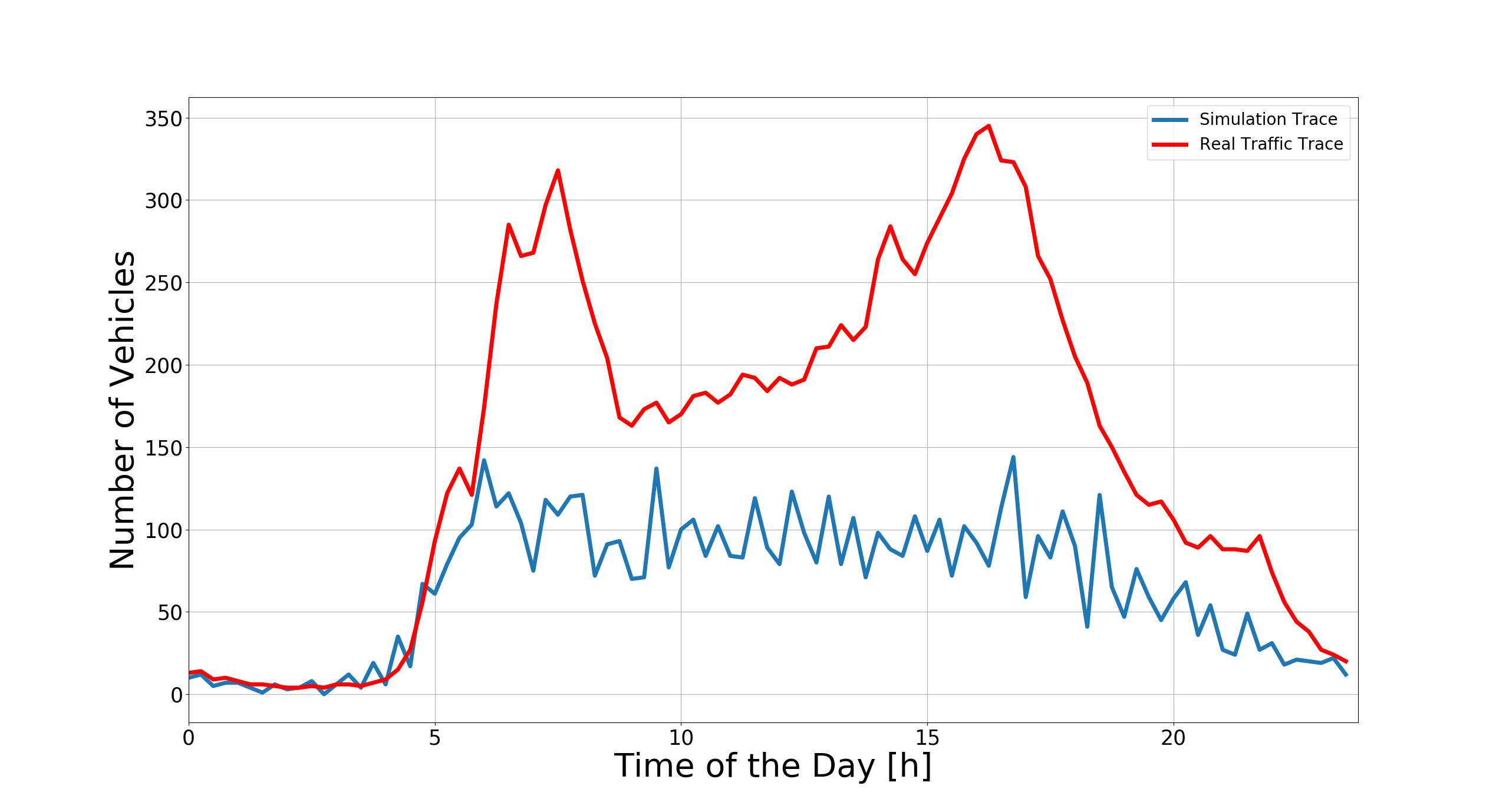}
        \caption{Worst Case}
        \label{fig:3150} 
\end{subfigure}
\caption{Best and Worst Crossings Evaluation}
\end{figure}

The best point is the junction between the arterial road Westliche Ringstrasse and the way Probierlweg. This junction is a three-way intersection, where approximately 18,414 vehicles drive daily. Four vehicle detectors were implemented in this junction. All detectors are placed on Westliche Ringstrasse, three in the north direction, and one detector in the south direction. Figure~\ref{fig:4210} shows the different behavior from the ST and RT, where it is observed that both traces are close with few mismatches periods.

The highest NRMSE has been observed on an intersection between the arterial roads N\"ordliche Ringstraße and the street Eckstallerstraße. Five detectors are thus in place. Figure~\ref{fig:3150} shows the behavior of this intersection, where it is possible to evaluate that the mismatch of both traces is relevant. The RT shows that in this crossing a large number of vehicles drive through during the day. Analyzing ST, it is observed that in the simulation this junction is not used as frequently by the drivers as one would expect.

\section{Conclusion and Future Work}
\label{sect:future-work}
This paper introduced InTAS, the Ingolstadt Traffic Scenario for SUMO. This traffic scenario is the first SUMO-based scenario using programs close to the deployed in the real traffic light system, and not only standard programs provided by SUMO. Traffic light programs' lengths and phases for twenty crossings were provided by the City of Ingolstadt and simulated into the scenario, where approximately 87,500 vehicles daily drive. InTAS represents the road network of Ingolstadt, due to the work to correct all the streets based on the information on-line available on satellite view from Google Maps. Traffic modeling took into consideration where people live and where they spend their daily activities, like work, school, and spare time. All these features establish an environment for simulations, seeking a real-world representation, and can cooperate with all kinds of vehicular simulations, e.g. C-ITS and VANETs. Furthermore, soon, this scenario will represent one of the first cities on Germany where the Car2X communication system is available~\cite{audi-vernetzt}. However, InTAS is the first realistic traffic scenario for SUMO analyzing a City with this feature. InTAS presents the public transport network, simulating 56 bus lines running over 1,620 daily trips and covering 880.6 km of routes length. A total of 21,756 buildings were inserted aiming to create an environment for network simulations, allowing effects such as signal-fading and shadow areas. 

Ingolstadt Scenario has been modeled and validated using real traffic information from 24 measurement points. A data-set was elaborated based on the information from each junction. This data-set computed the average number of vehicles driving through the crossing for an entire day. Detectors to count the vehicle's number are placed on all lanes from each intersection. An algorithm was implemented to reach the best \textit{device.rerouting.probability} value, comparing the simulation output and the vehicle's number from October of 2019. To validate the simulation output, virtual detectors were placed on the simulation as close as they are in the real world. The detectors' output was compared with the real data-set, based on November of 2019, creating traffic traces to be analyzed. The first analysis compared the total number of vehicles for all detectors, presenting an NRMSE of 0.33. Thereafter, the NRMSE for each intersection was evaluated, and the best and worst-case were deeper discussed.

The main issue faced on this research was the mismatch between the real and simulation-traces. This error might be due to, the traffic demand modeling method took into consideration the average values for the parameters \texttt{population} and \texttt{workPositions} for each of Ingolstadt's sub-area. When the average value is implemented, it implies particular errors. Hence, it is possible to model demographic data with more details, considering smaller regions inside the sub-areas and as this information is not available online, it is necessary to strengthen the relationship with the City of Ingolstadt to get this data. The other point is the time-base between demographic data and traffic data. Demographic data was published by the authorities considering the year 2018. On the other hand, real traffic used to validate InTAS were collected between August and December 2019. Even though the time between the demographic data and traffic is only eight months, it might change the traffic numbers and add errors. 

Furthermore, a fitness function for weighting SUMO parameters, the relevant VANETs features, and traffic realism could be implemented. This function can evaluate the impact on traffic realism considering all of SUMO's parameters and lead to a better fit, decreasing the error value. Among the solutions approaches to define the optimum value is Artificial Intelligence~\cite{concGA}, as Genetic Algorithm to find the best fit.

Following the tradition of SUMO, the InTAS scenario is freely available for the research community at \url{https://github.com/silaslobo/InTAS}.

\section*{Acknowledgments}
\label{sect:acks}

The authors would like to thank the Ingolstadt Verkehrsmanagement und Geoinformation Office branch of the City of Ingolstadt for providing real traffic information, AUDI AG for their contribution to this research, and Quentin Delooz and Anupama Hegde for their support. The authors look forward to continue interaction in the future.

\label{sect:bib}
\bibliographystyle{plain}
\bibliography{easychair}

\begin{thebibliography}{10}

\bibitem{audi-vernetzt}
Audi AG.
\newblock Audi vernetzt sich mit ampeln in deutschland.
\newblock [online], 2019.
\newblock
  \url{https://www.audi-mediacenter.com/de/pressemitteilungen/audi-vernetzt-sich-mit-ampeln-in-deutschland-11649},
  last viewed January 2020.

\bibitem{etsi2}
Thomas Biehle and Klaus Krumbiegel.
\newblock Triggering {Conditions} and {Data} {Quality} {Dangerous} {Situation}.
\newblock Technical Report Release 1.4.0, CAR 2 CAR Communication Consortium,
  09 2019.

\bibitem{etsi3}
Jan Buchholz.
\newblock Triggering {Conditions} and {Data} {Quality} {Exchange} of {IRCs}.
\newblock Technical Report Release 1.4.0, CAR 2 CAR Communication Consortium,
  09 2019.

\bibitem{LUST}
L.~{Codeca}, R.~{Frank}, and T.~{Engel}.
\newblock Luxembourg \textsc{SUMO} \textsc{T}raffic (\textsc{L}u\textsc{ST})
  \textsc{S}cenario: 24 hours of mobility for vehicular networking research.
\newblock In {\em 2015 IEEE Vehicular Networking Conference (VNC)}, pages 1--8,
  Dec 2015.

\bibitem{MOST}
L.~{Codeca} and Jérôme {HÄRRI}.
\newblock Monaco \textsc{SUMO} traffic (\textsc{M}o\textsc{ST}) scenario: A 3d
  mobility scenario for cooperative its.
\newblock In {\em SUMO 2018- Simulating Autonomous and Intermodal Transport
  Systems}, volume~2, pages 43--55, 2018.

\bibitem{dijkstra}
Edsger~W. Dijkstra.
\newblock A note on two problems in connexion with graphs.
\newblock {\em Numerische mathematik}, 1(1):269--271, 1959.

\bibitem{etsi1}
Sebastian Engel.
\newblock Triggering {Conditions} and {Data} {Quality} {Adverse} {Weather
  Conditions}.
\newblock Technical Report Release 1.4.0, CAR 2 CAR Communication Consortium,
  09 2019.

\bibitem{etsi4}
Sebastian Engel.
\newblock Triggering {Conditions} and {Data} {Quality} {Stationary} {Vehicle}
  {Warning}.
\newblock Technical Report Release 1.4.0, CAR 2 CAR Communication Consortium,
  09 2019.

\bibitem{verkehrsentwicklungsplan}
Amt für Verkehrsmanagement~und Geoinformation.
\newblock Verkehrsentwicklungsplan 2025.
\newblock [online], 2018.
\newblock
  \url{https://www.ingolstadt.de/media/custom/2789\_1089\_1.PDF?1540885166},
  last viewed December 2019.

\bibitem{Gawron}
Christian Gawron.
\newblock {An Iterative Algorithm to Determine the Dynamic User Equilibrium in
  a Traffic Simulation Model}.
\newblock In {\em Proceedings of the 3rd Industrial Simulation Conference 2005,
  EUROSIS-ETI}, pages 285--290, 1998.

\bibitem{google-maps}
Google.
\newblock Google maps.
\newblock [online].
\newblock
  \url{https://www.google.com/maps/@48.7621617,11.4349463,6820m/data=!3m1!1e3},
  last viewed June 2019.

\bibitem{subareas}
Stadt Ingolstadt.
\newblock Kartografie: Statistik und stadtforschung.
\newblock [online].
\newblock
  \url{https://www.ingolstadt.de/media/custom/2789_150_1_g.JPG?1501493618},
  last viewed January 2020.

\bibitem{verkehrspolitische}
Stadt Ingolstadt.
\newblock Verkehrspolitische ziele der stadt ingolstadt.
\newblock [online].
\newblock
  \url{https://www.ingolstadt.de/Rathaus/Verkehr/Verkehrsmanagement/Verkehrsplanung/index.php?La=1&object=tx,465.4793.1&kat=&kuo=2&sub=0},
  last viewed January 2020.

\bibitem{INGOLSTADT}
Stadt Ingolstadt.
\newblock Ingolstadt in zahlen 2018/2019.
\newblock [online], 2018.
\newblock \url{https://ingolstadt.de/media/custom/465\_1995\_1.PDF?1534841095},
  last viewed September 2019.

\bibitem{districts}
Stadt Ingolstadt.
\newblock Kleinräumige statistiken zum 31.12.2018.
\newblock [online], 2019.
\newblock
  \url{https://www.ingolstadt.de/media/custom/3052_1647_1.PDF?1563171021}, last
  viewed October 2019.

\bibitem{INVG}
INVG.
\newblock Ingolstädter verkehrsgesellschaft.
\newblock [online].
\newblock \url{https://www.invg.de/}, last viewed November 2019.

\bibitem{krauss}
S~Krauß.
\newblock {\em {Microscopic Modeling of Traffic Flow: Investigation of
  Collision Free Vehicle Dynamics}}.
\newblock PhD thesis, {DLR-Forschungsbericht}, 08 1998.

\bibitem{SUMO}
Pablo~Alvarez Lopez, Michael Behrisch, Laura Bieker-Walz, Jakob Erdmann,
  Yun-Pang Fl{\"o}tter{\"o}d, Robert Hilbrich, Leonhard L{\"u}cken, Johannes
  Rummel, Peter Wagner, and Evamarie Wie{\ss}ner.
\newblock {Microscopic Traffic Simulation using SUMO}.
\newblock In {\em The 21st IEEE International Conference on Intelligent
  Transportation Systems}, pages 2575--2582. IEEE, November 2018.

\bibitem{Ottawa}
D.~{McKenney} and T.~{White}.
\newblock Distributed and adaptive traffic signal control within a realistic
  traffic simulation.
\newblock In {\em Engineering Applications of Artificial Intelligence},
  volume~26, pages 574--583, January 2013.

\bibitem{environment}
R.~{Miucic}.
\newblock Introduction.
\newblock In {\em Connected Vehicles}, pages 1--10, 2019.

\bibitem{OpenROUTS3D}
S.~{Neumeier}, M.~{H\"opp}, and C.~{Facchi}.
\newblock Yet another driving simulator openrouts3d: The driving simulator for
  teleoperated driving.
\newblock In {\em 2019 IEEE International Conference on Connected Vehicles and
  Expo (ICCVE)}, pages 1--6, Graz, AUT, Nov 2019.

\bibitem{Car2XHIL}
C.~{Obermaier}, R.~{Riebl}, and C.~{Facchi}.
\newblock Fully reactive hardware-in-the-loop simulation for vanet devices.
\newblock In {\em 2018 21st International Conference on Intelligent
  Transportation Systems (ITSC)}, pages 3755--3760, Nov 2018.

\bibitem{Openstreetmap}
OpenStreetMap.
\newblock Open\textsc{S}treet\textsc{M}ap.
\newblock [online].
\newblock \url{https://openstreetmap.org}, last viewed June 2019.

\bibitem{gevas}
Gevas Humberg~\& Partner.
\newblock Integriertes verkehrsmodell datenerfassungskonzepte ingolstadt.
\newblock [online].
\newblock
  \url{https://www.gevas-ingenieure.de/pdf/REF_VP_Verkehrsmodell_Ingolstadt_IN-VMO02_neu.pdf},
  last viewed November 2019.

\bibitem{radioPropagationModel}
Muhammad~Ahsan Qureshi, Rafidah Md.~Noor, Azra Shamim, Shamshirband PhD, and
  Kim-Kwang~Raymond Choo.
\newblock A lightweight radio propagation model for vehicular communication in
  road tunnels.
\newblock {\em PLoS ONE}, 11, 03 2016.

\bibitem{VilaReal}
J.~{Soares}, C.~{Lobo}, Z.~{Vale}, and P.~B. {de Moura Oliveira}.
\newblock Realistic traffic scenarios using a census methodology: Vila real
  case study.
\newblock In {\em 2014 IEEE PES General Meeting | Conference Exposition}, pages
  1--5, July 2014.

\bibitem{SUMODUA}
SUMO.
\newblock Demand/dynamic user assignment.
\newblock [online], 2020.
\newblock \url{https://sumo.dlr.de/docs/Demand/Dynamic_User_Assignment.html},
  last viewed June 2020.

\bibitem{SUMOTimeLoss}
SUMO.
\newblock Simulation/output/tripinfo.
\newblock [online], 2020.
\newblock \url{https://sumo.dlr.de/docs/Simulation/Output/TripInfo.html}, last
  viewed June 2020.

\bibitem{concGA}
Agrani Swarnkar and Anil Swarnkar.
\newblock Artificial intelligence based optimization techniques: A review.
\newblock In Akhtar Kalam, Khaleequr~Rehman Niazi, Amit Soni, Shahbaz~Ahmed
  Siddiqui, and Ankit Mundra, editors, {\em Intelligent Computing Techniques
  for Smart Energy Systems}, pages 95--103, Singapore, 2020. Springer
  Singapore.

\bibitem{Buildings}
H.~{Tchouankem}, T.~{Zinchenko}, and H.~{Schumacher}.
\newblock Impact of buildings on vehicle-to-vehicle communication at urban
  intersections.
\newblock In {\em 2015 12th Annual IEEE Consumer Communications and Networking
  Conference (CCNC)}, pages 206--212, Jan 2015.

\bibitem{trafficlightimpact}
T.~{Tielert}, M.~{Killat}, H.~{Hartenstein}, R.~{Luz}, S.~{Hausberger}, and
  T.~{Benz}.
\newblock The impact of traffic-light-to-vehicle communication on fuel
  consumption and emissions.
\newblock In {\em 2010 Internet of Things (IOT)}, pages 1--8, Nov 2010.

\bibitem{TAPAS}
S.~{Uppoor} and M.~{Fiore}.
\newblock Large-scale urban vehicular mobility for networking research.
\newblock In {\em 2011 \textsc{IEEE} \textsc{V}ehicular \textsc{N}etworking
  \textsc{C}onference (\textsc{VNC})}, pages 62--69, Nov 2011.

\bibitem{trafficlight}
N.~{Wu}, D.~{Li}, and Y.~{Xi}.
\newblock Distributed weighted balanced control of traffic signals for urban
  traffic congestion.
\newblock {\em IEEE Transactions on Intelligent Transportation Systems},
  20(10):3710--3720, Oct 2019.

\end{thebibliography}


\end{document}